\documentclass[journal,10pt]{IEEEtran}
\makeatletter
\def\ps@headings{%
\def\@oddhead{\mbox{}\scriptsize\rightmark \hfil \thepage}%
\def\@ adversarynhead{\scriptsize\thepage \hfil \leftmark\mbox{}}%
\def\@oddfoot{}%
\def\@ adversarynfoot{}}
\makeatother
\ifCLASSINFOpdf
\else

\fi

\newcounter{problem}
\newcounter{save@equation}
\newcounter{save@problem}

\makeatletter

\usepackage{epsfig}
\usepackage{array}
\newcolumntype{L}[1]{>{\raggedright\let\newline\\\arraybackslash\hspace{0pt}}m{#1}}
\newcolumntype{C}[1]{>{\centering\let\newline\\\arraybackslash\hspace{0pt}}m{#1}}
\newcolumntype{R}[1]{>{\raggedleft\let\newline\\\arraybackslash\hspace{0pt}}m{#1}}
\usepackage[short]{optidef}
\usepackage{amsthm} 
\usepackage{mathrsfs}
\newcommand{\bc}{\begin{center}}
\newcommand{\ec}{\end{center}}
\newcommand{\be}{\begin{equation}}
\newcommand{\ee}{\end{equation}}
\usepackage{flexisym}
\usepackage{algorithmicx}
\usepackage{algorithm}
\usepackage{siunitx}
\usepackage{soul}
\usepackage{algpseudocode}

\newcommand{\bnu}{\begin{enumerate}}
\newcommand{\enu}{\end{enumerate}}
\usepackage{cite}
\usepackage{url}
\usepackage[utf8]{inputenc}
\usepackage[T1]{fontenc}
\usepackage{amsmath}
\usepackage{amsfonts}
\usepackage{amssymb}
\usepackage[version=4]{mhchem}
\usepackage{stmaryrd}
\usepackage{graphicx}
\usepackage[export]{adjustbox}
\graphicspath{ {./images/} }
\usepackage{breqn}
\usepackage{lipsum}
\usepackage{mathtools}
\usepackage{caption}
\usepackage{float}
\usepackage{subcaption}
\usepackage{soul}
\usepackage{blindtext, graphicx}
\usepackage{cuted}
\usepackage{color}

\usepackage{amsfonts}
\usepackage{multicol}
\usepackage{lipsum}
\usepackage{cuted}
\newtheoremstyle{case}{}{}{}{}{}{:}{ }{}

\begin{document}
\title{Optimal Microcontroller Usage in Reconfigurable Intelligent Surface: Batteryless IoT Systems Case Study}
\author{Shakil Ahmed, \IEEEmembership{Member,~IEEE}, Ahmed E. Kamal, \IEEEmembership{Fellow,~IEEE} and Mohamed Y. Selim, \IEEEmembership{Senior Member,~IEEE}
\vspace*{-0.75 cm}
\thanks{
Corresponding Author: S. Ahmed (shakil@iastate.edu)\\
Shakil Ahmed, Ahmed E. Kamal, and Mohamed Y. Selim are with the Department of Electrical and Computer Engineering, Iowa State University, Ames, Iowa, USA (email: \{shakil, kamal, myoussef\}@iastate.edu). \\
A preliminary version of this paper was presented at 2022 IEEE Local Computer Networks (LCN) Conference.}
}
{}

\maketitle

\begin{abstract}
To effectively enhance wireless communications in internet of things (IoT) systems using reconfigurable intelligent surfaces (RISs), it is crucial to have proper control over programmable passive elements for reflection angle manipulation and active elements that amplify and reflect signals in the desired direction. However, incorporating a high number of RIS elements requires an adequate number of microcontrollers, which results in increased system complexity and cost.
This paper proposes a solution to improve the efficiency of RISs in enhancing wireless communications in IoT systems by investigating the trade-off between the number of RIS elements and microcontrollers.
Our solution is a novel approach and differentiating factor from other works in the literature,  named "Module," in which one microcontroller controls a single module. 
We define the modules as either active or passive, each containing an optimal number of active or passive elements.
We determine the optimal number of elements to establish the size of a module using a non-linear energy harvesting model where a batteryless IoT (b-IoT) sensor harvests energy from a nearby base station (BS) radio frequency (RF) signals.
Using the optimal size of a module, we find the optimal number of modules, i.e., the optimal number of microcontrollers for efficient RIS panel controlling. With the harvested energy from the BS RF signals, the b-IoT sensor transmits data from the BS to other IoT sensors, which enforces an information causality constraint in RIS module-assisted IoT systems. 
We formulate an optimization problem to minimize the energy consumption of the RIS modules while satisfying the non-linear energy harvesting model and information causality constraint.
The formulated non-convex mixed-integer non-linear energy minimization problem is solved using iterative algorithms. 
Simulation results show that the RIS module-assisted energy harvesting enhances the performance of the IoT systems by around 100\% compared to benchmark models with no RIS panel.
\end{abstract}

\section{Introduction}
\IEEEPARstart{A}{s} IoT systems continue to grow and converge with smart sensing, computation, and communication, it becomes crucial to deploy intelligent environments to provide massive connectivity \cite{Int_0}. This is because mobile data traffic is projected to increase by 30\% annually from 2018 to 2024, with each sensor data usage ranging from 6 to 23 gigabytes per month \cite{Eric_1}.
IoT systems integrate the physical and digital world in various wireless domains, including e-healthcare systems, autonomous vehicles, smart cities, smart homes, smart cities, and intelligent transportation systems. They rely on state-of-the-art information and employ wireless technologies for efficient and effective operations. The potential of IoT technology is demonstrated by the estimated deployment of billions of IoT sensors soon \cite{Chpt_Mod_Ref_IoT4}.

IoT sensors are frequently portable, which presents a challenge in maintaining their operations due to the need for a steady power supply. The recent exponential growth in batteryless IoT (b-IoT) sensors further complicates the issue \cite{number_bIoT}. Deploying these sensors in remote or inaccessible locations without nearby power sources makes powering them even more difficult \cite{Ref_IoT9}. Therefore, the continued operation of b-IoT sensors depends heavily on a reliable energy source. However, traditional power supplies such as batteries and solar panels require significant installation and maintenance costs. The increasing number of b-IoT sensors creates a growing demand for wireless power transfer as an alternative. Great visionary Nikola Tesla was among the pioneers of wireless energy transfer, developing the concepts of energy transmission via the air medium and energy transformation into usable direct current (DC) \cite{Ref_IoT10}. His foresight led to one of the first achievements in understanding and developing state-of-the-art power supply techniques, including energy harvesting.

Among various techniques for energy harvesting, IoT sensors can collect energy directly from external radio frequency (RF) sources, such as neighboring base stations (BSs), which is a promising solution to power the sensors \cite{Ref_EH2}. However, IoT sensors consume considerable energy due to continuous services, which may shorten their lifetime in an energy-constrained scenario. Harvesting energy from BS RF signals can address this issue by converting RF energy to electrical energy \cite{Ref_EH1}. However, an enhanced channel gain is essential to overcome the complexity of the energy harvesting process and ensure improved performance. The reconfigurable intelligent surface (RIS) technology can address the challenges by adjusting the direction of incident signals in a controlled manner and enhancing the gains between nodes. RISs improve energy harvesting and enhance the communication performance between nodes, making them an essential component for next-generation technology. Therefore, using RISs to ensure efficient channel gains for wireless communications is an attractive approach for forthcoming technologies.

The RIS is a flat structure composed of individually configured elements designed to control electromagnetic waves dynamically. These elements induce a specific phase change in reflected signals, allowing for more efficient wireless propagation. RIS elements can be passive or active, with passive elements reflecting the incident signals and active elements amplifying them to the receiving nodes. This technology enhances energy efficiency in IoT systems by directing and strengthening transmit signals to the destination nodes. Our prior research in \cite{Ref_RIS_Shakil, Ref_RIS_Shakil1} has determined the optimal number of RIS elements, both active and passive, to maximize data rate and energy harvesting for b-IoT systems. These studies found that adding active elements can reduce the number of total RIS elements while improving system performance. Therefore, RIS panels offer significant advantages over conventional wireless communication methods.

A large number of RIS elements may be deployed to compensate for the double-fading attenuation \cite{Ref_RIS3,Ref_RIS4,Ref_RIS5,Ref_RIS6,Ref_RIS7, Ref_RIS8}. This deployment results in a larger RIS panel surface area with considerable power consumption due to the internal circuits, causing significant overhead, such as higher complexity and expensive maintenance. Due to the complexity of RIS panel properties, optimizing RIS parameters, such as phase shift, reflection angle, amplitude, etc., is also difficult. Another major challenge to controlling the RIS panels intelligently still exists, which needs adequate research to find a trade-off between the number of intelligent controllers, such as microcontrollers, and the RIS panel size. This control allows the RIS to manipulate the electromagnetic waves that pass through it to achieve the desired results, such as reflecting, focusing, or scattering signals. Without optimal number of microcontrollers, it would be impossible to control the behavior of the individual elements that make up the RIS, and the full potential of the surface would not be realized. Therefore, using microcontrollers is a critical aspect of RIS technology, and their precise control is essential for the efficient functioning of RISs. Unfortunately, trivial attention is given to microcontroller-based efficient RIS controlling. Finding an optimal trade-off guarantees a minimum number of microcontrollers with less operational complexity, cost, and RIS energy consumption. Therefore, architectural and operational challenges must be addressed to enjoy the full benefits of the RIS panel. This paper aims to find the optimal number of microcontrollers to control the bias voltage and achieve optimal performance without compromising the system's performance, such as optimal energy harvesting and data transmissions.

\subsection{Related Works and Motivation}

Researchers in the wireless community have extensively studied various critical aspects of IoT systems, including energy harvesting, maximizing bit transmission, and the utilization of RIS panels \cite{Ref_RW_1,Ref_RW_2,Ref_RW_3,Ref_RW_5,Ref_RW_6,Ref_RW_9,Ref_RW_11,Ref_RIS14, Ref_RW_13,Ref_RW_14,Ref_RW_15,Ref_RW_1..,Ref_RW_17,Ref_RW_18,Ref_RW_20,Ref_RW_7,Ref_RW_8,Ref_RW_19}. Specifically, \cite{Ref_RW_1,Ref_RW_2,Ref_RW_3,Ref_RW_5,Ref_RW_6,Ref_RW_9,Ref_RW_11,Ref_RIS14} discussed IoT systems, while \cite{Ref_RW_13,Ref_RW_14,Ref_RW_15,Ref_RW_1..,Ref_RW_17,Ref_RW_18,Ref_RW_20,Ref_RW_7,Ref_RW_8,Ref_RW_19} emphasized the significance and potential deployment of RIS panels.
For instance, \cite{Ref_RW_1} proposed a novel framework to optimize the achievable rate for IoT data transmission from the base station (BS) to the sensors. In \cite{Ref_RW_2}, the authors suggested that harvesting energy from RF sources is the best option for low-power IoT sensors. Furthermore, \cite{Ref_RW_3} designed an energy harvesting model for low-power IoT sensors to explore environmental monitoring systems. In \cite{Ref_RW_5}, a new concept was introduced to power IoT sensors by harvesting energy from nearby smart devices like smartphones. Lastly, \cite{Ref_RW_6} investigated IoT systems where energy is shared among the sensors.

The authors in \cite{Ref_RW_9} designed a cognitive radio network that maximizes resource allocation, including efficient information and energy transfer. To achieve this, they considered a secondary user with an unspecified backlog of traffic and an error-free sensing function. They calculated the energy region of a wireless system and looked at optimizing the region among the nodes in an opportunistic mobile network, considering a transmitter, a decoder, and an energy harvesting device. In \cite{Ref_RW_11}, the authors expanded on this model to maximize system throughput for secondary users using a stochastic approach to consider transmission power. They assumed that users could harvest energy for each other if they were close. Additionally, \cite{Ref_RIS14} focused on studying the features of the BS RF signals and microcontrollers to configure the status of the RF systems for b-IoT sensors that use harvested energy to convert external energy sources, such as BS RF signals, into DC energy.
The RIS panel has been proposed in various wireless network domains, such as IoT sensors, cognitive radio networks, device-to-device (D2D) communications, unmanned aerial vehicles (UAVs), etc.

The possible deployment of the RIS panels integrated into IoT systems significantly enhances the system performance \cite{Ref_RW_13,Ref_RW_14,Ref_RW_15,Ref_RW_1..,Ref_RW_17,Ref_RW_18,Ref_RW_7,Ref_RW_8,Ref_RW_20,Ref_RW_19}.
For example, with the help of the RIS panel, a link between a BS and a user was considered to maximize the throughput by increasing the elements of the RIS panel in \cite{Ref_RW_13}.
They also focused on maximizing energy efficiency in RIS-assisted wireless networks.
The RIS-aided future wireless network was proposed in \cite{Ref_RW_14} for indoor and outdoor scenarios.
The authors considered a single RIS panel to maximize the system performance matrices.
The authors in \cite{Ref_RW_15} studied relay-based wireless networks with the help of the RIS panel.
In their proposed networks, the RIS panel enhanced the link between transmitter and receiver in the first hop, while the other hop was enhanced by relay.
The required channel state information estimation was considered in \cite{Ref_RW_1..} by embedding RIS active sensors.
The authors in \cite{Ref_RW_17} used the Fisher-Snedecor \textit{F} model to estimate the channel in the RIS panel-assisted wireless networks.
The mean square error for RIS panel-assisted IoT systems was investigated in \cite{Ref_RW_18} to minimize the mean square error in their model.
The authors in \cite{Ref_RW_20} investigated a rate maximization problem with the help of the RIS panel and UAV.
While energy harvesting from RF sources has been widely studied, most research is limited to decoding information with a non-linear energy harvesting model \cite{Ref_RW_7, Ref_RW_8}.
A framework for RIS panel-assisted low-power IoT systems was studied to investigate the back-scatter communications in \cite{Ref_RW_19}.
Unfortunately, none of the works in \cite{Ref_RW_1,Ref_RW_2,Ref_RW_3,Ref_RW_5,Ref_RW_6,Ref_RW_9,Ref_RW_11,Ref_RIS14,Ref_RW_13,Ref_RW_14,Ref_RW_15,Ref_RW_1..,Ref_RW_17,Ref_RW_18,Ref_RW_7,Ref_RW_8,Ref_RW_20,Ref_RW_19} considered RIS modules and the optimal number of microcontrollers to enhance the performance of their models.

\subsection{Contributions}
In our research, we focus on improving the performance of RIS module-assisted wireless networks for b-IoT sensors while reducing the number of microcontrollers required. Our key contributions are as follows:

\begin{itemize}

\item \textbf{Designing a module's size:} 
We optimize energy harvesting and data transmission by determining the optimal number of RIS elements used for communications between IoT sensors through harvesting energy from BS RF signals. To overcome potential fading or shadowing in networks, we incorporate RIS panels consisting of active and passive elements. Furthermore, to achieve efficient RIS controlling with less complexity, we determine the optimal number of RIS elements and use it later to define the optimal module size.

\item \textbf{RIS active and passive modules:}
The main contribution of our work is developing a novel approach for controlling RIS panels that we call a "Module." This approach involves defining the optimal size of active and passive RIS modules using a non-linear energy harvesting model and then determining the optimal number of modules needed for efficient RIS panel operation.
The information causality constraint ensures that the information is forwarded from the BS to the next IoT sensor. The goal is to determine the optimal number of modules while reducing the total energy consumption of the RIS modules.

\item \textbf{Trade-off between the microcontrollers and RIS size:}
We establish a trade-off between the number of microcontrollers and RIS elements and show that each microcontroller can control a single RIS module for efficient control with less hardware complexity and cost.
The RIS modules are, therefore, controlled by the equivalent number of microcontrollers, which provide efficient control for these modules with less hardware complexity and co

\item \textbf{Efficient algorithm and performance evaluation:} 
We address the challenging non-convex mixed integer non-linear programming (MINLP) problem of minimizing RIS modules' energy consumption using an efficient iterative algorithm and show that RIS module-assisted systems outperform conventional wireless networks. Our findings also demonstrate that using a larger number of RIS elements leads to significant improvement by reducing the required number of microcontrollers.
\end{itemize}

\subsection{Notations and Paper Organization}
\begin{figure*}[!h]
\centering
\includegraphics[width=6.8in]{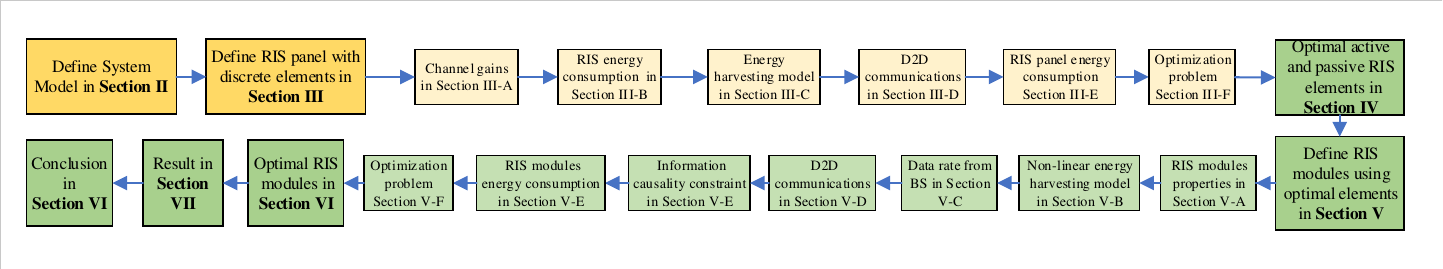}
\caption{Paper organization}
\label{Flow}
\end{figure*}

We distinguish scalar, vector, and matrix with the variables $a$, $\mathbf{a}$, and $\mathbf{A}$, respectively.
We also define $[.]^*$, $[.]^{\dagger}$, and $[.]^f$ as optimal values, conjugate transpose, and feasible points, respectively.
Fig.~\ref{Flow} shows the flow diagram of the paper organization.
We define the proposed system model in Section~\ref{SM}. 
The detailed description of RIS with discrete elements assisted networks and the proposed solution of finding optimal elements are illustrated in Sections~\ref{RDE} and \ref{OAPE}, respectively.  
Similarly, the proposed model with RIS modules and the proposed iterative algorithm to achieve optimal modules are explained in Sections~\ref{RMM} and \ref{OAPM}, respectively.
While the results are shown in Section~\ref{RA12}, we conclude in Section~\ref{End}.

\section{System Model} \label{SM}

\begin{table}[]
\caption{List of mathematical symbols}
\begin{tabular}{llll}
\hline
\multicolumn{1}{|c|}{\textbf{Symbol}}                     & \multicolumn{1}{c|}{\textbf{Description}}      & \multicolumn{1}{c|}{\textbf{Symbol}}              & \multicolumn{1}{c|}{\textbf{Description}}             \\ \hline
\multicolumn{4}{|c|}{\textbf{Constants}}                          \\ \hline
\multicolumn{1}{|l|}{$f_{\{.\}}$}                         & \multicolumn{1}{l|}{Constant}       & \multicolumn{1}{l|}{$T(i^{'})$}          & \multicolumn{1}{l|}{Energy harvesting time}  \\ \hline
\multicolumn{1}{|l|}{$\lambda$}                   & \multicolumn{1}{l|}{Wavelength}       & \multicolumn{1}{l|}{$T(i^{''})$}         & \multicolumn{1}{l|}{D2D communications time} \\ \hline
\multicolumn{1}{|l|}{$\!\!f_{\{.\}}\!\!\!$} & \multicolumn{1}{l|}{Constant}         & \multicolumn{1}{l|}{$N$}             & \multicolumn{1}{l|}{RIS elements}            \\ \hline
\multicolumn{1}{|l|}{$\sigma^2$}                  & \multicolumn{1}{l|}{Noise}            & \multicolumn{1}{l|}{$p_{sc}$}            & \multicolumn{1}{l|}{Switch/control power}    \\ \hline
\multicolumn{1}{|l|}{$p_{dc}$}                    & \multicolumn{1}{l|}{DC power}         & \multicolumn{1}{l|}{$a_m$}               & \multicolumn{1}{l|}{Amplification factor}    \\ \hline
\multicolumn{1}{|l|}{$W_{\{.\}}$}               & \multicolumn{1}{l|}{Bandwidth}        & \multicolumn{1}{l|}{$\delta_{\{.\}}$}    & \multicolumn{1}{l|}{Path loss component}     \\ \hline
\multicolumn{1}{|l|}{${h_{\{.\}}}$}            & \multicolumn{1}{l|}{Gains}    & \multicolumn{1}{l|}{$\zeta$}             & \multicolumn{1}{l|}{Efficiency factor}       \\ \hline
\multicolumn{1}{|l|}{$\mathbf{h_{rd}}$}           & \multicolumn{1}{l|}{RIS-$D$ gain}     & \multicolumn{1}{l|}{$\alpha_i, \beta_i$} & \multicolumn{1}{l|}{Reflection coefficient}  \\ \hline
\multicolumn{1}{|l|}{$\mathbf{h_{sr}}$}           & \multicolumn{1}{l|}{$S$-RIS gain}     & \multicolumn{1}{l|}{$s$}       & \multicolumn{1}{l|}{Unit power signals}      \\ \hline
\multicolumn{1}{|l|}{$\Upsilon_{\{.\}}$}                         & \multicolumn{1}{l|}{Rician factor}    & \multicolumn{1}{l|}{$Y,y_{\{.\}}$}       & \multicolumn{1}{l|}{Constants}               \\ \hline
\multicolumn{1}{|l|}{$h_{bs}$}                    & \multicolumn{1}{l|}{BS-$S$ gain}      & \multicolumn{1}{l|}{$T(i)$}         & \multicolumn{1}{l|}{Time duration for frame, $i$}                \\ \hline
\multicolumn{1}{|l|}{$\mathbf{h_{br}}$}           & \multicolumn{1}{l|}{BS-RIS gain}      & \multicolumn{1}{l|}{$h_{sd}$}            & \multicolumn{1}{l|}{$S$-$D$ gain}            \\ \hline
\multicolumn{1}{|l|}{$\mathbf{h_{rs}}$}           & \multicolumn{1}{l|}{RIS-$S$ gain}      & \multicolumn{1}{l|}{$\mathbf{z_{\{.\}}}$}            & \multicolumn{1}{l|}{RIS thermal noise}            \\ \hline
\multicolumn{1}{|l|}{$d_{\{.\}}$}           & \multicolumn{1}{l|}{Distance}      & \multicolumn{1}{l|}{$p_{out,m}$}            & \multicolumn{1}{l|}{Active elements output power}            \\ \hline
\multicolumn{1}{|l|}{$\sigma^2_{\{.\}}$}           & \multicolumn{1}{l|}{Noise}      & \multicolumn{1}{l|}{$p_{in,m}$}            & \multicolumn{1}{l|}{Active elements input power}            \\ \hline
\multicolumn{4}{|c|}{\textbf{Variables}} \\ \hline
\multicolumn{1}{|l|}{$m$}                   & \multicolumn{1}{l|}{Active elements}  & \multicolumn{1}{l|}{$p_{ct}^{b}(i)$}          & \multicolumn{1}{l|}{Active module power at slot   $i$}     \\ \hline
\multicolumn{1}{|l|}{$k$}                     & \multicolumn{1}{l|}{Passive elements} & \multicolumn{1}{l|}{$p_{ss}^{b}(i)$}          & \multicolumn{1}{l|}{Passive module power at slot   $i$}    \\ \hline
\multicolumn{1}{|l|}{$p_b$}                       & \multicolumn{1}{l|}{BS power}         & \multicolumn{1}{l|}{$e$}                 & \multicolumn{1}{l|}{Linear harvested energy} \\ \hline
\multicolumn{1}{|l|}{$\boldsymbol{\Theta,\Phi}$}                       & \multicolumn{1}{l|}{RIS properties}   & \multicolumn{1}{l|}{$p_r^l$}             & \multicolumn{1}{l|}{Linear harvested power}  \\ \hline
\multicolumn{1}{|l|}{$e_{nl}^{b}$}                 & \multicolumn{1}{l|}{Harvested energy} & \multicolumn{1}{l|}{$p_{ct}$}            & \multicolumn{1}{l|}{Active element power}    \\ \hline
\multicolumn{1}{|l|}{$[.]^f$}                       & \multicolumn{1}{l|}{Feasible point}    & \multicolumn{1}{l|}{$p_{ss}$}            & \multicolumn{1}{l|}{Passive element power}   \\ \hline
\multicolumn{1}{|l|}{$e_r^{b}$}                      & \multicolumn{1}{l|}{Module energy}   & \multicolumn{1}{l|}{$K_p$}           & \multicolumn{1}{l|}{Number of passive modules}       \\ \hline
\multicolumn{1}{|l|}{$p_{nl}^{b}$}                      & \multicolumn{1}{l|}{Harvested power}   & \multicolumn{1}{l|}{$M_a$}           & \multicolumn{1}{l|}{Number of active modules}       \\ \hline
\multicolumn{1}{|l|}{$[.]^*$}                      & \multicolumn{1}{l|}{Optimal value}   & \multicolumn{1}{l|}{$e_{nl},e^b_{nl}$}           & \multicolumn{1}{l|}{Harvested energy}       \\ \hline
\end{tabular}
\end{table}

We investigate IoT systems consisting of BS and IoT sensors, $S$ and $D$, communicating in RIS-assisted wireless networks.
The BS and IoT sensors are equipped with a single antenna.
An RIS panel is placed in the network to enhance the system's performance.
We envisage two separate scenarios to minimize the energy consumption of the RIS panel, such as the RIS panel equipped with 1) active and passive elements (shown in Fig.~\ref{System_Model1}),
2) active and passive modules (shown in Fig.~\ref{System_Model2}).
We define active and passive modules that comprise a number of active and passive elements, respectively.
Based on far-field communications theory, we assume that all the elements in a module are controlled in the same way for phase shifts and amplitude.
The BS is considered to have single antenna elements, while each sensor has a single antenna.

\subsection{RIS Panel with Discrete Elements} \label{RIS_s}
We design a wireless system where \textcolor{black}{the total time is a repetition of the time frame.}
We define   $i$ as the frame number while  $i^{'}$, and $i^{''}$  are the slots of frame   $i$.
The duration of frame, $i$, is $T(i)$ which is divided two slots with duration $T(i^{'})$ and $T(i^{''})$.
We also assume the coherence time is much longer than $T(i)$; hence, the channel conditionals within $T(i)$ are quasistatic.
The RIS panel consists of $N$ discrete elements with $m$ active and $k$ passive elements, where $m+k \leq N$.
The RIS panel assists the IoT sensor, $S$, to harvest energy from the BS RF signals in $T(i^{'})$ and also in D2D communications from $S$ to $D$ in $T(i^{"})$.
We use a non-linear model for energy harvesting from the BS RF signals.
We focus on finding the optimal number of elements given by $m^*$ and $k^*$, respectively, to minimize the RIS panel energy consumption during $T(i)$.
Note that $m^*$ and $k^*$ are used later to design the size of active and passive modules, respectively.

\subsection{RIS Panel with Modules}
We re-investigate the time-slotted wireless network from Section~\ref{RIS_s} with the following extensions,
\textit{(1)} Each RIS module is controlled by a single microcontroller, providing efficient control for these modules with less controlling complexity, power consumption, and cost.
We define $M_a$ as the number of active modules, where each module is equipped with $m^*$ number of elements.
$K_p$ is the number of passive modules, where each module is equipped with the number of $k^*$ elements.
\textit{(2)} 
IoT sensor, $S$, harvests energy from the BS RF signals and receives information from the BS during $T(i^{'})$.
\textit{(3)} $S$ communicates to $D$ during $T(i^{''})$.
The model enforces the information causality constraint to forward the information received from the BS to another IoT sensor, $D$.
Finally, we minimize overall energy consumed by RIS modules during $T(i)$.

\section{RIS Panel with Discrete Elements} \label{RDE}
\begin{figure}[!h]
\centering
\includegraphics[width=3.35in]{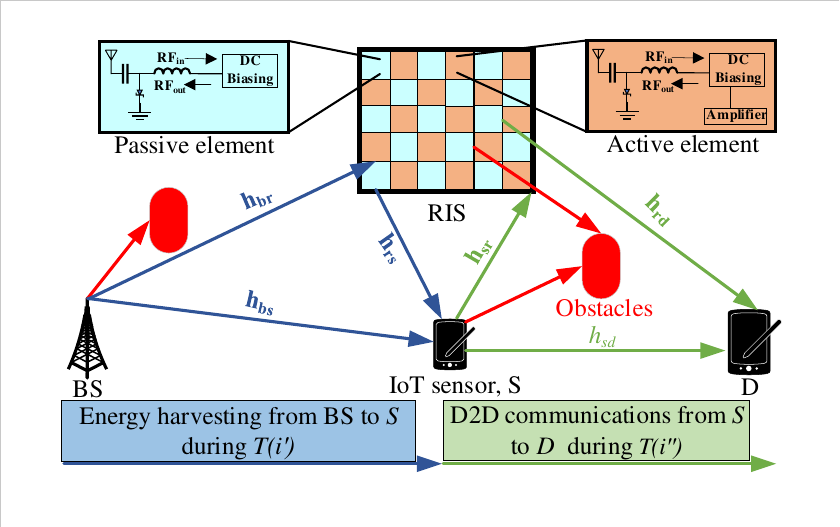}
\caption{System model with RIS elements}
\label{System_Model1}
\end{figure}
\subsection{Channel Gains} \label{NoBlock_gain}
Without loss of generality, we assume that the channel gains are the same during each transmission frame with noise variance, $\sigma^2$.
The gains from $S$ to $D$, BS to $S$,  BS to RIS, RIS to $S$, $S$ to RIS, and RIS to $D$ are defined as $h_{sd}$, $h_{bs}$,  $\mathbf {h_{br}} \in \mathbb{C}^{N\times 1}$, $\mathbf {h_{rs}} \in \mathbb{C}^{1\times N}$, $\mathbf {h_{sr}} \in \mathbb{C}^{N\times 1}$, and $\mathbf {h_{rd}} \in \mathbb{C}^{1\times N}$, respectively.
The channel gains of the communications links from BS to $S$ and $S$ to $D$ are modeled using Rician fading and, respectively, are given as follows: 
\be
\begin{aligned}
&h_{bs}=\frac {\sqrt {\beta _{bs}^{-1}}}{\sqrt {\Upsilon_e+1}}(\sqrt {\Upsilon_e} h_{bs}^{los}+h_{bs}^{nlos})
\end{aligned}
\ee
\be
\begin{aligned}
h_{sd}=\frac {\sqrt {\beta _{sd}^{-1}}}{\sqrt {\Upsilon_s+1}}(\sqrt {\Upsilon_s} h_{sd}^{los}+h_{sd}^{nlos})
\end{aligned}
\ee
where $\beta_{bs}=\frac{16 \pi^2 d_1^{\delta_{1}}}{\lambda^2}$ and $\beta_{sd}=\frac{16 \pi^2 d_2^{\delta_{2}}}{\lambda^2}$.
$\lambda$ is the wavelength.
$\delta_1^p$ and $\delta_2^p$ are the path loss exponent.
$d_1$ and $d_2$ are communicating node distance.
The elements of $h_{bs}^{nlos}$, $h_{bs}^{los}$, $h_{sd}^{nlos}$, and $h_{sd}^{los}$ are independently and identically distributed according to $\mathcal{CN}(0,1)$.
$ \Upsilon_{\{e,s\}} \in [0,+\infty) $ is Rician factor.

\subsection{RIS Elements}
A wavelength is larger than a single RIS element, which can be as small as $\frac{\lambda}{8} \times \frac{\lambda}{8}$. The RIS elements scatter the signal with approximately equal gain toward the receiving nodes. As shown in Fig.~\ref{System_Model1}, the RIS panel consists of $N$ discrete elements. $m$ and $k$ represent the number of active and passive elements, respectively, with $m + k \leq N$. 
We use $m(i^{'}) \leq m$ and $k(i^{'}) \leq k$ during $T(i^{'})$ while $m(i^{''}) \leq m$  and $k(i^{''}) \leq k$ are used during $T(i^{''})$.

\subsubsection{Energy harvesting from BS RF signals} 
$N \times N$ diagonal matrix that describes the amplitude and phase shift introduced by the RIS panel elements during the energy harvesting phase is given as follows:
\be
\begin{aligned}
\!\!\! \boldsymbol{\Theta} = & \mathrm {diag} (\alpha_1 e^{j \theta _{1}}, \alpha_2 e^{j \theta_2}, .., \alpha_k e^{j \theta _{k}}, \\
& \alpha_{k+1} e^{j \theta _{k+1}}, \alpha_{k+2} e^{j \theta_{k+2}}, .., \alpha_{N} e^{j \theta _{N}} )
\end{aligned}
\ee
where the fixed amplitude reflection coefficient is defined as $\alpha$.
We limit $\alpha_j \in [0,1] $, where $i=1,2,...,k$ for passive elements and $\alpha_j \geq 1 $, where $j=k+1,k+2,...,n$ for active elements.
The phase-shift of the RIS elements is defined as $\theta$.
We define $\{ \theta_1,\theta_2,...,\theta_{k} \}$ for passive elements and $\{ \theta_{k+1},\theta_{k+2},...,\theta_{N} \}$ for active elements.

\subsubsection{Data transmission}
The IoT sensor, $S$, communicates with another IoT sensor, $D$, using a D2D communications protocol. During the data transmission phase, the $N \times N$ diagonal matrix that describes the amplitude and phase shift introduced by the RIS panel elements is given as follows:
\be
\begin{aligned}
\boldsymbol{\Phi} = & \mathrm {diag} (\beta_1 e^{j \phi_{1}}, \beta_2 e^{j \phi_2}, .., \beta_k e^{j \phi_{k}}, \\
& \beta_{k+1} e^{j \phi_{k+1}}, \beta_{k+2} e^{j \phi_{k+2}}, .., \beta_{N} e^{j \phi_{N}} )
\end{aligned}
\ee
where the fixed amplitude reflection coefficient is defined as $\beta$.
We limit $\beta_j \in [0,1] $, where $i=1,2,...,k$ for passive elements and $\beta_j \geq 1 $, where $j=k+1,k+2,...,n$ for active elements.
The phase-shift of the RIS elements is defined as $\phi$.
We define $\{ \phi_1,\phi_2,...,\phi_{k} \}$ for passive elements and $\{ \phi_{k+1},\phi_{k+2},...,\phi_{N} \}$ for active elements, respectively.

\subsection{Non-Linear Energy Harvesting Model}
To accurately design the size of a module, we utilize a non-linear energy harvesting model that reflects real-world scenarios. 
The received signal from BS at $S$ is:
\be
\begin{aligned}
s_{bs}(i^{'})=&\underbrace { \sqrt{p_{b}}h_{bs} s }_{\text {direct link}}+\underbrace {\mathbf {h_{rs}}\boldsymbol{\Theta }\left ({ \sqrt{p_{b}}\mathbf {h_{br}} s +\mathbf {z}_{2} }\right)}_{{\text {RIS-aided link}}}+{z}_{1} \\=&\sqrt{p_{b}}\left ({ {h}_{bs}+ \mathbf {h_{rs}}\boldsymbol{\Theta }\mathbf {h_{br}}}\right)s+ \mathbf {h_{rs}}\boldsymbol{\Theta }\mathbf {z}_{2} + {z}_{1}
\end{aligned}
\ee
where $s$ is the unit-power information signal.
$\mathbf{z_2}\in \mathbb{C}^{N\times 1}$ defines the generated thermal noise of the RIS panel during the energy harvesting from the BS RF signals with a distribution of $\mathcal{CN}(0,\sigma_2^2)$.
$z_1$ is the additive Gaussian channel noise with a  distribution of $\mathcal{CN}(0,\sigma_1^2)$.
For a given $\boldsymbol{\Theta}$, we achieve the capacity of the additive white Gaussian noise channel with the channel gains as follows:
\be
\begin{aligned}
\mathbf {h_{rs}} \boldsymbol{\Theta} \mathbf {h_{rb}^{\dagger}}= \sum_{n=1}^N \alpha_n e^{j \theta_n} [\mathbf {h_{rs}}]_n [\mathbf {h_{rb}^{\dagger}}]_n
\end{aligned}
\ee
where $N$ is the number of the RIS elements.
The maximum rate is achieved if $\theta_n$ is optimally selected to maximize the sum of $\mathbf {h_{rs}} \boldsymbol{\Theta} \mathbf {h_{rb}^{\dagger}}$. $\theta_n$ is expressed as follows:
\be
\begin{aligned}
\theta_n= \arg(\mathbf{h_{bs}})- \arg([\mathbf {h_{rs}}]_n [\mathbf {h_{rb}^{\dagger}}]_n)
\end{aligned}
\ee

Non-linear energy harvested from BS RF signals during $T(i^{''})$ is expressed as follows \cite{Ref_NEHModel}:
\be
\begin{aligned}
e_{nl}(i^')=\frac{YT(i^{'})}{1+\exp(-y_1(p_{nl}(i^')-y_2))}
\end{aligned}
\ee
where 
\be
\begin{aligned}
p_{nl}(i^')= \zeta p_b \frac{(h_{bs}+{ (a_mm(i^')+k(i^')){h_1}})^{2}}{  \sigma_1^2}
\end{aligned}
\ee

In this definition, ${h_1}=[\mathbf {h_{rs}}]_n [\mathbf {h_{rb}^{\dagger}}]_n  $ and $\sigma_1^2$ is the noise.
$a_m$ is the amplification factor.
The active and passive elements used during $T(i^')$ are represented by $m(i^')$ and $k(i^')$, respectively.
$\zeta \in [0,1]$ is the power split coefficient that indicates the portion of power assigned to the energy harvesting unit.
$Y$, $y_1$, and $y_2$ define the various parameters required for the wireless energy harvesting model, where $y_1$ and $y_2$ capture the effects of resistance, capacitance, and circuit sensitivity.

\subsection{D2D Communications}
The received signal at sensor, $D$ during $T(i^{''})$ is expressed as follows:
\be
\begin{aligned}
s_{bs}(i^{'})=&\underbrace { \sqrt{\frac{e_{nl}(i^{'})}{T(i^{''})}}h_{sd} s }_{\text {direct link}}+\underbrace {\mathbf {h_{rd}}\boldsymbol{\Phi }\left ({ \sqrt{\frac{e_{nl}(i^{'})}{T(i^{''})}} \mathbf {h_{sr}}s +\mathbf {z}_{2} }\right)}_{{\text {RIS-aided link}}}+{z}_{1} \\=&\sqrt{\frac{e_{nl}(i^{'})}{T(i^{''})}} \left ({ {h}_{sd}+ \mathbf {h_{rd}}\boldsymbol{\Phi }\mathbf {h_{sr}}}\right)s+ \mathbf {h_{rd}}\boldsymbol{\Phi}\mathbf {z}_{2} + {z}_{1}
\end{aligned}
\ee

For given $\boldsymbol{\Phi }$, the gains are defined as follows:
\be
\begin{aligned}
\mathbf {h_{rd}} \boldsymbol{\Phi } \mathbf {h_{rs}^{\dagger}}= \sum_{n=1}^N \beta_n e^{j \phi_n} [\mathbf {h_{rd}}]_n [\mathbf {h_{rs}^{\dagger}}]_n
\end{aligned}
\ee
where $\mathbf {h_{rs}^{\dagger}}$ is conjugate transpose.
Using $\mathbf {h}_{\mathrm {rd}} \boldsymbol{\Phi} \mathbf {h}_{\mathrm {sr}}^{\dagger}$, the maximum rate is achieved if $\phi_n$ is optimally selected to maximize the sum to be the same phase of $h_{sd}$. $\phi_n$ is, therefore, expressed as follows:
\be
\begin{aligned}
\phi_n=\arg(h_{sd}) - \arg([\mathbf {h_{rd}}]_n [\mathbf {h_{rs}^{\dagger}}]_n)
\end{aligned}
\ee

We calculate the rate during $T(i^{''})$ using harvested energy from BS RF signals as follows:
\be
\begin{aligned}
\label{Rate_D2D}
&R_{d}(i^{''})=\! \! \! \! \max _{\phi_{1},\ldots,\phi_{N}} \! \! \! \! W_1 \! \log _{2} \! \!\left ({\! \!1\!  + \! \frac {\frac{e_{nl}(i^{'})}{T(i^{''})} ( h_{sd}\! + \! \mathbf {h}_{\mathrm {rd}}^{ { \mathrm {}}} \boldsymbol{\Phi} \mathbf {h}_{\mathrm {rs}}^\dagger)^{2}}{\sigma_e^2} \! \!}\right) \\
&= W_1 \! \log _{2} \! \! \left ({ \! \! 1 \! + \! \frac { \frac{e_{nl}(i^{'})}{T(i^{''})} ({h_{sd} \! + \! (a_m m(i^{'' \! })+k(i^{'' \! })) {h_2} })^{2}}{\sigma_e^2} } \! \right)
\end{aligned}
\ee
where the carrier bandwidth is defined as $W_1$. 
We define 
$ {h_2}= [\mathbf {h_{rd}}]_n [\mathbf {h_{rs}^{\dagger}}]_n$ and $\sigma_e^2=\sigma_2^2||\mathbf{h_{rd}} \mathbf{\Phi} ||^2+\sigma_1^{2}$. 

\subsection{RIS Elements Energy Consumption} \label{Sec_III_E}
This section focuses on calculating the energy consumption of the RIS elements by considering both the energy harvesting and data transmission phases. The energy consumption in the passive elements is primarily due to the circuits related to the switch and control. On the other hand, active elements have a higher energy consumption because of the power consumed by the active loads and power amplification, in addition to the energy consumed by the switch and control circuits. 

\subsubsection{Passive elements power consumption}
Recall that $k$ is the number of RIS passive elements.
We express the RIS power consumption due to passive elements as follows:
\be
\begin{aligned}
\label{eq_ac_powe}
p_{ss}(i)=k(j) p_{sc}~\text{for}~j\in \{i^{'},i^{''}\}
\end{aligned}
\ee
where $p_{sc}$ is the energy consumption of each element, considering switch and control circuits.

\subsubsection{Active elements power consumption}
The focus of analyzing the active elements' energy consumption is on the DC circuit loss and the output power. 
The RIS active elements amplify reflected signals with negative resistance tunnel diodes and a DC polarization source, leading to power loss in the DC circuit. 
The output power encompasses the energy consumption by the switch and the control circuit.

To get a reasonably tractable and accurate power consumption model, the power amplifier model from \cite{joung2014survey} is used to characterize the power consumption of the reflection amplifier for active elements. 
The power consumed by $m$ number of identical active elements during energy harvesting is expressed as follows:
\be
\begin{aligned}
\label{p_ct_1}
p_{ct}= m(p_{ss}+p_{dc}) +   \text{OPD}(p_{out,m})
\end{aligned}
\ee

The RIS active elements' power consumption is analyzed with a focus on the power dissipation caused by the DC circuit loss and the output power. The output power includes the energy consumed by both the switch and control circuit, represented by $p_{ss}$, and the DC biasing power consumption, represented by $p_{dc}$, which can be expressed as the sum of these two, i.e., $(p_{ss}+p_{dc})$. The output power dependent term, $\text{OPD}(p_{out,m})$, is determined by the output power, $p_{out,m}$ for $m$ number of active elements. This term can be modeled linearly through empirical expression as follows \cite{9377648}:
\be
\begin{aligned}
\label{p_ct_2}
\text{OPD} (p_{out,m})=\frac{p_{out,m}}{v}=\frac{a_m^2p_{in,m}}{v}
\end{aligned}
\ee

The output power for $m$ number of elements is related to the incident signal power $p_{in,m}$ through the amplifier efficiency, $v$. This relationship can be expressed as follows:
\be
\begin{aligned}
\label{p_ct_3}
P_{in,m} =  ||\mathbf{h_{{br}}}||^2p_b+\sigma_2^2
\end{aligned}
\ee

The reflection amplifier operates in its linear region, and the output power grows proportionally to the input power. This holds for tunnel diode-based reflection amplifiers when the input power is within its operational range \cite{lee2016rf}. The BS-RIS channel attenuation may weaken the transmitted signal, allowing the reflection amplifier to achieve a high amplification gain with low output power. 
The output amplification power is generally a minor component of the overall power consumption, compared to the DC power consumption \cite{amato2018tunneling}. As such, most studies only account for DC power consumption when estimating the power consumption of the reflection amplifier. The active elements of the RIS provide signal amplification without the need for complex and power-hungry RF components, reducing power consumption. 
The power consumption of reflection amplifiers was reduced to $\mu$W.
The system benefits from the improved signal combination at the receiver and increased power amplification with the help of RIS active elements.

\textit{Power consumption during energy harvesting:}
Using (\ref{p_ct_1})~-~(\ref{p_ct_3}) we formulate the power consumption expression as follows:
\be
\begin{aligned}
\label{eq_ac_powe}
p_{ct}(i^{'}) \! = \! m(i^{'}) (p_{sc} \! + \! p_{dc}) \! + \frac{{p_{out_1}}}{v}
\end{aligned}
\ee
where the output power of the active RIS, represented by $p_{out_1}$,  is defined as $p_{out_1} = a_m^2 p_b||\mathbf{h_{br}}||^2 + \sigma_2^2$. 

\textit{Power consumption during D2D communications:}
Similarly, the active elements power consumption during D2D communications phase with the help of harvested energy is expressed as follows:
\be
\begin{aligned}
\label{eq_ac_powe2}
\!\!\!\!p_{ct}(i^{''}) \! = \! m(i^{''}) (p_{sc} \! + \! p_{dc}) \! + \frac{p_{out_2}}{v}
\end{aligned}
\ee
where $p_{out_2} = a_m^2 \frac{e_{nl}(i^{'})}{T(i^{''})}|| \mathbf{h_{sr}}||^2+\sigma_2^2$.

\subsubsection{RIS panel energy consumption}
Overall RIS panel energy consumption during $T(i)$ is expressed as follows:
\be
\begin{aligned}
\label{eq_tot_powe..}
e_{r}(i)=   \sum_{j \in \{{i^{'},i^{''}}\}}T(j) \bigg [ p_{ss}(j)+p_{ct}(j) \bigg ]
\end{aligned}
\ee


\subsection{Optimization Problem}
This section formulates an optimization problem that minimizes the RIS elements' energy consumption during time frame $i$ subject to a non-linear energy harvesting model and data transmission.
We formulate the optimization problem, named as $\textbf{\text{P.1}}$, as follows:
\begin{subequations}\label{eq_op_1}
\begin{align}\label{eq_ob_1}\
&{\textbf{P.1}~\mathop {\min }\limits_{m, k,p_b,T(i^{'}),T(i^{''})}
}\ {\sum_{j \in \{{i^{'},i^{''}}\}}T(j) \bigg [ p_{ct}(j)+p_{ct}(j) \bigg ] } \\
&\text{s.t.}\ e_m \leq \frac{YT(i^{'})}{1+\exp(-y_1(p_{nl}(i^')-y_2))} \label{eq_opt1_c1}\\
& \! \! \!R_d^t \!\leq \! W_1 \! \log _{2} \! \! \left ({ \! \! 1 \! + \! \frac { \frac{e_{nl}(i^{'})}{T(i^{''})} ({h_{sd} \! + \! (a_m m(i^{'' \! })\!+\!k(i^{'' \! })) {h_2} })^{2}}{\sigma_e^2}}  \! \right)\!\!\! \label{eq_opt1_c2}\\
& T(i^{'})+T(i^{''}) = T(i) \label{eq_opt1_c3}\\
& m+k \leq  N \label{eq_opt1_c3.}\\
& m(j) \leq m,~~k(j) \leq k~\text{for}~ j \in \{i^{'},i^{''}\} \label{eq_opt1_c3.3}
\end{align}
\end{subequations}

The objective function minimizes the RIS elements' energy consumption over both time slots and is described in (\ref{eq_ob_1}).
Moreover, (\ref{eq_opt1_c1}) defines that $S$ harvests energy from BS RF signals above a threshold, $e_m$.
In (\ref{eq_opt1_c2}), we define minimizing data transmission rate from $S$ to $D$.
The number of active and passive elements used in two-time slots are shown in (\ref{eq_opt1_c3.})~-~(\ref{eq_opt1_c3.3}).
Note that (\ref{eq_op_1}) is not only a non-convex problem due to coupling of optimizing variables in $p_{ct}(i)$ of (\ref{eq_ob_1})~-~(\ref{eq_opt1_c2}) but also MINLP that makes  (\ref{eq_op_1}) difficult to solve optimally for any practically sized problem. 
In order to address the non-convex MINLP, a heuristic solution approach is presented in Section~\ref{OAPE}.

\section{Optimal Number of RIS Elements} \label{OAPE}
The objective function in (\ref{eq_op_1}) describes a non-convex optimization problem, which is challenging to solve due to the interaction of the coupling variables in the objective function (\ref{eq_ob_1}), energy harvesting constraint (\ref{eq_opt1_c1}), and data transmission constraint (\ref{eq_opt1_c2}). The optimization problem is transformed into three equivalent subproblems to make the problem more tractable.
The first subproblem optimizes the RIS panel elements, given the BS transmit power and their corresponding time slots.
The second subproblem uses the optimal RIS elements to optimize the time slots.
The final subproblem optimizes the BS transmit power using the optimal elements and time slots.
Each of these subproblems is investigated in turn in the following subsections.

\subsection{Optimal RIS elements, $m^*$ and $k^*$}
To provide an efficient and practical solution to (\ref{eq_op_1}) and optimize the number of active and passive elements, $m$ and $k$, respectively,
we reformulate (\ref{eq_op_1}) as \textbf{P.1A}:
\begin{subequations}\label{eq_op_3}
\begin{align}\label{eq_ob_2.}\
&{\textbf{P.1A}~ \mathop {\!\!\!\!\min\!\!\!\!\!\! }\limits_{m, k}
}\ {\sum_{j \in \{{i^{'},i^{''}}\}}T(j) \bigg [ p_{ct}(j)+p_{ct}(j) \bigg ]} \\
&\text{s.t.}\ e_m \!\! \leq \! \zeta T(i^') p_{b} \frac{\left({h_{bs}+ (a_m m(i^')+k(i^')){h_1}}\right)^{2}}{{\sigma_1^{2}}} \label{eq_opt2_c1}\\
& 2^{\frac{R_{d}^t}{W_1}} \!\! \leq 1 \! + \! \frac{\frac{e_{nl}(i^{'})}{T(i^{''})} ({ h_{sd}+( a_m m(i^{''\!}) \! + \! k(i^{''\!}) ) {h_2} })^{2}}{{\sigma_e^2}} \! \!\! \label{eq_opt2_c6}\\
& (\ref{eq_opt1_c3.})~-~(\ref{eq_opt1_c3.3}) \nonumber
\end{align}
\end{subequations}

After some mathematical manipulations, we reformulate the non-convex energy harvesting constraint in (\ref{eq_opt2_c1}) by taking the exponential term out as follows:
\be
\begin{aligned}
\label{eq_sub_1}
&  f_1  \geq  h_{bs}^2+2h_{bs}h_1(a_mm(i^{'})+k(i^{'}))+h_1^2
 a_m^2m^f(i^{'}) \\  
 & m(i^{'}) +2a_mh_1m(i^{'})k^f(i^{'})+h_1^2k(i^{'})k^f(i^{'})
\end{aligned}
\ee
where 
\be
\begin{aligned}
f_1=\sigma_{1}^2\frac{\bigg(y_2-\frac{\ln(\frac{YT(i^')}{e_m}-1)}{y_1} \bigg) }{\zeta p_b}
\end{aligned}
\ee

In (\ref{eq_opt2_c6}), we reformulate to become convex as follows: 
\be
\begin{aligned}
\label{eq_sub_3}
f_2 \leq e_{nl}(i^{'})({ h_{sd}+( a_m m(i^{''\!}) \! + \! k(i^{''\!}) ) {h_2} })^{2}
\end{aligned}
\ee
where $f_2=\sigma_e^2 T(i^{''}) ( 2^{\frac{R_{d}^t}{W_1}}- 1) $.
We expand the RHS of (\ref{eq_sub_3}) as follows:
\be
\begin{aligned}
&  \underbrace{e_{nl}(i^{'})h_{sd}^2}_{\text{Item 1}}+\underbrace{2{h_2}h_{sd} a_me_{nl}(i^') m(i^{''\!})}_{\text{Item 2}} + \! \underbrace{2{h_2}h_{sd}e_{nl}(i^')k(i^{''\!}) }_{\text{Item 3}} + \\
& ~~~~~\underbrace{{h_2}^2a_m^2 e_{nl}(i^{'})m^2(i^{''})}_{\text{Item 4}} +\underbrace{ 2 {h_2}^2 a_m e_{nl}(i^{'})m(i^{''})k(i^{''}) }_{\text{Item 5}}+ \\
& ~~~~~~~~~~~~~~~~~~~\underbrace{ {h_2}^2 e_{nl}(i^{'})k^2(i^{''})}_{\text{Item 6}} \nonumber
\end{aligned}
\ee

Items 1-6 are non-convex due to the quadratic terms and coupling variables.
We decouple and linearize non-linear terms at feasible points.

\textbf{Item 1:}
We define item 1 as $t_1$. Using (\ref{eq_sub_1}), we reformulate $t_1$ as follows:
\be
\begin{aligned}
\label{Eq_l1}
&t_1 \!= \! f_3 \bigg( h_{bs}+2 h_{bs} {h_1} \bigg( \! a_m m (i^')+ k(i^') \bigg)
\\
& + {h_1^2} \left({ \! a_m m(i^') + k(i^') }\right) \left({ a_m m^f(i^') + k^f(i^')}\right)\bigg)
\end{aligned}
\ee
where
\be
\begin{aligned}
f_3=h_{sd}^2 \zeta T(i^') p_{b}
\end{aligned}
\ee

\textbf{Item 2:}
We reformulate the item 2, defined as $t_2$, using (\ref{eq_sub_1}) as follows:
\be
\begin{aligned}
\label{Eq_nl1}
& t_2 \!=\! f_4 \bigg( h_{bs}m(i^{''}) +2 h_{bs} {h_1} \left({ a_m m^f (i^') \!+ \! k^f(i^')}\right)\!\! \\
&m(i^{''}\!) \!\!+\!\! \left({ \! a_m m(i^'\!) \! + \! k(i^'\!)\! }\right)\!\! \left({ a_m m^f\!(i^'\!) \!+\! k^f\!(i^'\!)\!}\right) \!m(i^{''}\!) {h_1^2} \!\bigg) \!\!
\end{aligned}
\ee
where
\be
\begin{aligned}
f_4=2 {h_2} h_{sd}a_m\zeta T(i^') p_{b}
\end{aligned}
\ee

\textbf{Item 3:}
We tackle item 3, defined as $t_3$, as follows:
\be
\begin{aligned}
\label{Eq_nl2}
& t_3 \!=\! f_5 \bigg( h_{bs}k(i^{''}) +2 h_{bs} {h_1} \left({ a_m m^f (i^') \!+ \! k^f(i^')}\right)\!\! \\
&k(i^{''}\!) \!\!+\!\! \left({ \! a_m m(i^'\!) \! + \! k(i^'\!)\! }\right)\!\! \left({ a_m m^f\!(i^') \!+\! k^f(i^')\!}\right) \!k(i^{''}\!) {h_1}\!^2 \!\bigg) \!\!\!\!
\end{aligned}
\ee
where
\be
\begin{aligned}
f_5=2 {h_2} h_{sd}\zeta T(i^') p_{b}
\end{aligned}
\ee

\textbf{Item 4:}
Similarly, item 4, is defined as $t_4$ and expressed as follows:
\be
\begin{aligned}
\label{Eq_nl3}
&\!\!\!t_4\!=\!f_6  (h_{bs}m(i^{''})\!+\!2h_{bs}{h_1}m^f(i^{''\!})(a_mm(i^{'\!})\!+\!k(i^{'\!})) \!\!\! \\
& +{h_1^2} m(i^{''})(a_mm^f(i^{'\!})\!+\!k^f(i^{'\!})) (a_mm(i^{'\!})\!+\!k(i^{'\!})))
\end{aligned}
\ee
where
\be
\begin{aligned}
f_6= {h_2}^2 a_m^2 \zeta T(i^') p_b m^f(i^{''})
\end{aligned}
\ee

\textbf{Item 5:}
Item 5, defined as $t_5$, is expressed as follows:
\be
\begin{aligned}
\label{Eq_nl4}
& \!\!t_5\!= \!f_7 (k(i^{''}\!)h_{bs}+2h_{bs}{h_1} k^f(i^{''}\!)(a_mm(i^'\!)+k(i^'\!)) \!\!\!\\
&\!\!\!\!+{h_1^2} k^f(i^{''})(a_mm(i^')+k(i^'))(a_mm^f(i^')+k^f(i^')))
\end{aligned}
\ee
where
\be
\begin{aligned}
f_7= 2{h_2}^2 a_m \zeta T(i^') p_bm^f(i^{''})
\end{aligned}
\ee

\textbf{Item 6:}
We apply a similar approach to define the last item.
\be
\begin{aligned}
\label{Eq_nl5f}
&\!\!\!t_6\!=\!f_8 (h_{bs}k(i^{''})\!+\!2h_{bs}{h_1}k^f(i^{''\!})(a_mm(i^{'\!})\!+\!k(i^{'\!})\!) \!\!\! \\
& +{h_1^2} k(i^{''})(a_mm^f(i^{'\!})\!+\!k^f(i^{'\!})) (a_mm(i^{'\!})\!+\!k(i^{'\!})))
\end{aligned}
\ee
where
\be
\begin{aligned}
f_8={h_2}^2\zeta T(i^') p_b k^f(i^{''})
\end{aligned}
\ee

Using (\ref{Eq_l1})~-~(\ref{Eq_nl5f}), we reformulate (\ref{eq_sub_3}) as follows:
\be
\begin{aligned}
\label{Ref_eq21}
f_2 \leq t_1  + \! t_2 \! + \! t_3 \! + \! t_4 \! + \! t_4 \! + \! t_6 \! \! \!
\end{aligned}
\ee

Since we decouple items 1-6, we conclude that (\ref{Ref_eq21}) is to be convex.
Note that the objective function in (\ref{eq_ob_2.}) is not still convex due to $p_{ct}(i)$. We reformulate $p_{ct}(i)$ from (\ref{eq_ob_1}) using (\ref{eq_ac_powe2}) and (\ref{eq_sub_1}) as follows:
\be
\begin{aligned}
\label{eq_RIS}
p_{ct}^f(i^{''}) \! = \! m(i^{''}) (p_{sc} \! + \! p_{dc}) \! + \frac{a_m^2 \frac{e_{nl}^f(i^{'})}{T(i^{''})}|| \mathbf{h_{sr}}||^2+\sigma_2^2}{v}
\end{aligned}
\ee
where
\be
\begin{aligned}
& e_{nl}^f(i^')={YT(i^{'})}/\bigg( 1+\exp \bigg (-y_1 \bigg ( \zeta p_b \\
& \frac{(h_{bs}+{ (a_mm^f(i^')+k^f(i^')){h_1}})^{2}}{  \sigma_1^2} -y_2\bigg )\bigg ) \bigg)
\end{aligned}
\ee

The reformulated problem from (\ref{eq_op_3}) is expressed as follows:
\begin{align} \label{eq_op_3.}
& \underset{m^*, k^*}{\text{min~}~} {T(i^{'})p_{ct}(i^{'}) + T(i^{''})p_{ct}^f(i^{''})+\!\!\!\sum_{j \in \{i^{'},i^{''}\}}\!\!\!\!\!\!T(j)   k(j) p_{sc} } \\
& \text{s.t.}\ (\ref{eq_opt1_c3.})~-~(\ref{eq_opt1_c3.3}), (\ref{eq_sub_1}), (\ref{Ref_eq21}),(\ref{eq_RIS}) \nonumber
\end{align}

Note that (\ref{eq_op_3.}) is a convex problem. Feasible points, $m^f$ and $k^f$ are initialized to some reasonable random values.
The feasible points are updated at every iteration until it the optimal solutions, $m^*$ and $k^*$.

\subsection{Optimal time slots, $T^*(i^{'})$ and $T^*(i^{''})$}
To achieve $T^*(i^{'})$ and $T^*(i^{''})$, we reformulate (\ref{eq_op_1}) as follows:
\begin{subequations}\label{eq_sop_1}
\begin{align}\label{eq_sob_1}\
&{\textbf{P.1B}~\mathop {\min }\limits_{T(i^{'}),T(i^{''})}
}\ {\sum_{j \in \{{i^{'},i^{''}}\}}T(j) \bigg [ p_{ct}(j)+p_{ct}(j) \bigg ] }\\
&\text{s.t.}\ e_m \leq \frac{YT(i^{'})}{1+\exp(-y_1(p_{nl}(i^')-y_2))} \label{eq_sopt1_c1}\\
& \! \! \!R_d^t \!\leq \! W_1 \! \log _{2} \! \! \left ({ \! \! 1 \! + \! \frac { \frac{e_{nl}(i^{'})}{T(i^{''})} ({h_{sd} \! + \! (a_m m(i^{'' \! })\!+\!k(i^{'' \! })) {h_2} })^{2}}{\sigma_e^2}}  \! \right)\!\!\! \label{eq_sopt1_c2}\\
& (\ref{eq_opt1_c3}) \nonumber
\end{align}
\end{subequations}

Note that (\ref{eq_sopt1_c1}) is convex since no optimizing variables exist inside the exponential term. 
We approximate (\ref{eq_sopt1_c2})  to be convex as follows:
\be
\begin{aligned}
\label{Eq_nls2_1}
(2^{\frac{R^t_d}{W_1}}-1)\sigma_e^2T(i^{''}) \leq  \frac{YT(i^{'})}{1+\exp(-y_1(p_{nl}(i^')-y_2))}
\end{aligned}
\ee

Now (\ref{eq_sob_1}) is approximated as follows:
\begin{align} \label{eq_sop_1s}
& \underset{T^*(i^{'}),T^*(i^{''})}{\text{min}} {\sum_{j \in \{{i^{'},i^{''}}\}}T(j) \bigg [ p_{ct}(j)+p_{ct}(j) \bigg ] }   \\
&\text{s.t.}\ (\ref{eq_opt1_c3}), (\ref{eq_sopt1_c1}), (\ref{Eq_nls2_1}) \nonumber
\end{align}

\subsection{Optimal transmit power, $p_b^*$}
To achieve $p_b^*$, we reformulate (\ref{eq_op_1}) as follows:
\begin{subequations}\label{eq_sop_3}
\begin{align}\label{eq_sob_3}\
&{\textbf{P.1C}~\mathop {\min }\limits_{p_b}
}\ {\sum_{j \in \{{i^{'},i^{''}}\}}T(j) \bigg [ p_{ct}(j)+p_{ct}(j) \bigg ] }\\
&\text{s.t.}\ e_m \leq \frac{YT(i^{'})}{1+\exp(-y_1(p_{nl}(i^')-y_2))} \label{eq_sopt3_c1}\\
& \! \! \!R_d^t \!\leq \! W_1 \! \log _{2} \! \! \left ({ \! \! 1 \! + \! \frac { \frac{e_{nl}(i^{'})}{T(i^{''})} ({h_{sd} \! + \! (a_m m(i^{'' \! })\!+\!k(i^{'' \! })) {h_2} })^{2}}{\sigma_e^2}}  \! \right)\!\!\! \label{eq_sopt3_c2}
\end{align}
\end{subequations}

Due to the exponential term, it is not straightforward to make (\ref{eq_sopt3_c1}) convex. We simplify it as follows:
\be
\begin{aligned}
\label{Eq_nls3_1}
\!\!\!\frac{\ln\!\!\bigg(\!\! \frac{YT(i^{'})}{e_m}\!-1\!\!\bigg)}{y_2} \geq - \frac{\zeta p_b(h_{bs}+{ (a_mm(i^')+k(i^')){h_1}})^{2}}{\sigma_1^2}
\end{aligned}
\ee

We deal (\ref{eq_sopt3_c2}) to make it convex. We simplify it as follows:
\be
\begin{aligned}
\label{Eq_nls3_2}
\!\!y_1y_2-\ln \!\! \bigg(\!\frac{YT(i^{'})}{f_9} \!\bigg)\! \leq\! \frac{\zeta p_b (h_{bs}\!+\!{ (a_mm(i^')\!+\!k(i^')){h_1}})^{2}}{\sigma_1^2}\!\!
\end{aligned}
\ee

where 
\be
\begin{aligned}
f_{9}=\frac{\bigg( 2^{R^t_d}-1\bigg) \sigma_e^2}{T(i^{''})({h_{sd} \! + \! (a_m m(i^{'' \! })\!+\!k(i^{'' \! })) {h_2} })^{2}}
\end{aligned}
\ee

$p_{ct}(i^{"})$ from the objective function in (\ref{eq_sob_3}) is not convex yet due to the harvested energy. We use feasible points and approximate to be the convex problem. 

In order to make (\ref{eq_sob_3}), we approximate (\ref{eq_ac_powe2}) to be convex as follows:
\be
\begin{aligned}
\label{eq_RISsdsd}
p_{ct}^f(i^{''}) \! = \! m(i^{''}) (p_{sc} \! + \! p_{dc}) \! + \frac{a_m^2 \frac{e_{nl}^f(i^{'})}{T(i^{''})}|| \mathbf{h_{sr}}||^2+\sigma_2^2}{v}
\end{aligned}
\ee
where
\be
\begin{aligned}
& e_{nl}^f(i^')={YT(i^{'})}/\bigg ( 1+\exp \bigg (-y_1\bigg ( \zeta p_b^f \\
& \frac{(h_{bs}+{ (a_mm(i^')+k(i^')){h_1}})^{2}}{  \sigma_1^2} -y_2\bigg )\bigg )\bigg )
\end{aligned}
\ee

Now (\ref{eq_sop_3}) is approximated to convex as follows:
\begin{align} \label{eq_sop_s3}
& \underset{p_b^*}{\text{min~~}} T(i^{'})p_{ct}(i^{'})+ T(i^{''})p_{ct}^f(i^{''})+\!\!\!\!\!\!\!\sum_{j \in \{{i^{'},i^{''}}\}}\!\!\!\!\!\!T(j)  p_{ss}(j)   \\
&\text{s.t.}\ (\ref{Eq_nls3_1}), (\ref{Eq_nls3_2})\nonumber
\end{align}

\subsection{Proposed Algorithm}
We summarize the proposed solution in an algorithm as follows:
\begin{algorithm}
\caption{}
\label{alg:algorithm_sum}
\begin{algorithmic}[1]
\State $\bold{Initialization:}$ $m^f$, $k^f$, $T(i)$, $p_b^f$, and $\epsilon$ to some reasonable random values 
\State $\bold {Optimization}$:
\Repeat
\State Find the optimal RIS  active and passive elements, $m^*$ and $k^*$ using (\ref{eq_op_3.})
\State Set $m^f=m^*$ and $k^f=k^*$ 
\State Find the optimal time slots, $T^*(i^{'})$ and $T^*(i^{''})$ using (\ref{eq_sop_1s}), $m^*$ and $k^*$
\State Find the optimal BS transmit power, $p_b^*$ using (\ref{eq_sop_s3}), $T(i^{'})$, $T(i^{''})$, $m^*$ and $k^*$
\State Set $p_b^f=p_b^*$ 
\Until{Find the optimal value, such as $m^*$, $k^*$, $T^*(i^{'})$, $T^*(i^{''})$, and $p_b^*$ such that $|x(l)|-|x(l-1)|< \epsilon$, where $x=m^*,k^*,T^*(i^{'}),T^*(i^{''}),p_b^*$}
\end{algorithmic}
\end{algorithm}

\section{RIS Panel With Modules} \label{RMM}
\begin{figure}
\centering
\includegraphics[width=3.35in]{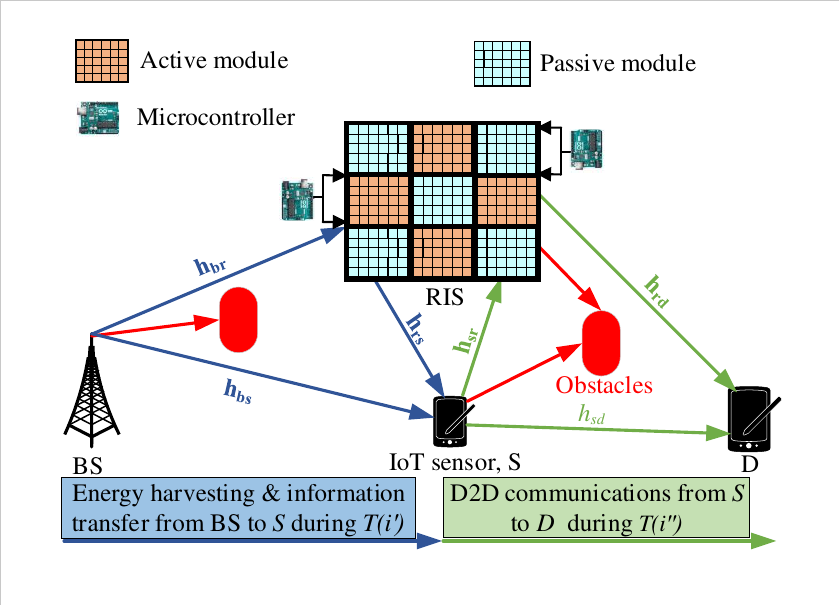}
\caption{System model with multiple RIS panel modules}
\label{System_Model2}
\end{figure}

The size of active and passive modules in the b-IoT system is determined in Section~\ref{RDE} using $m^*$ and $k^*$, respectively. This section aims to optimize the number of RIS modules while considering a non-linear energy harvesting model and information causality constraint and using the channel model from Section~\ref{NoBlock_gain}.
Each frame, denoted as   $i$, is divided into two slots, $i^{'}$ and $i^{''}$, with duration $T(i^{'})$ and $T(i^{''})$, respectively, such that $T(i) = T(i^{'}) + T(i^{''})$.
As depicted in Fig.~\ref{System_Model2}, during $T(i^{'})$, the b-IoT sensor $S$ harvests energy from the BS RF signals and receives information with the help of RIS modules. Using the harvested energy and RIS modules, $S$ then communicates to $D$ and forwards the received information from the BS.

\subsection{RIS Modules Properties}
This subsection outlines the size definitions for the active and passive modules, which consist of $m^*$ and $k^*$ elements, respectively. The RIS updates its configuration states from one to the next by receiving external commands via either a detached microcontroller architecture such as a field-programmable gate array (FPGA) or an integrated microcontroller architecture like a network of communicating chips/modules \cite{Ref_Micro}. FPGA-based architectures are typically bulky and consume a significant amount of power.
On the other hand, the integrated architecture uses a network of communicating chips/modules to read the state of the modules and adjust the bias voltages using impedance-adjusting semiconductors. These circuits receive, interpret, and apply the commands while exhibiting static and dynamic power consumption \cite{Ref_A.C.T}. 

The authors in \cite{Ref_CL} suggested that these integrated microcontrollers could work autonomously by harnessing energy from nano-networking advancements and using asynchronous logic, eliminating the need for complex and power-consuming clock distributions.
This control method, using integrated architecture, provides amplitude measurements of reflection coefficients at a given frequency and is more efficient than the near-field model. This design improves performance while reducing hardware complexity and cost by reducing the number of microcontrollers needed to control multiple RIS modules. 
The passive modules reflect incident signals and active modules amplify them, leading to additional energy consumption related to DC circuit power dissipation.

The RIS panel uses $M_a$ active modules and $K_p$ passive modules.
For a total number of modules, $(M_a+K_p)$, the RIS panel has the following number of elements used in energy harvesting and data transmission phases during the first and second-time slots, respectively, as follows:
\be
\begin{aligned}
M_am^*+ K_pk^* \leq N
\end{aligned}
\ee
where $N$ is the maximum number of elements used to design the active and the passive modules.
For ease of calculations, we redefine the following expression for the rest of the paper as follows:

\be
\begin{aligned}
L(j)= m^* \sum_{m=1}^{M_a(j)}a_m +K_p(j)k^*~\text{for}~j \in \{i^{'},i^{''}\}
\end{aligned}
\ee

It is important to note that this redefinition only applies within the context of the paper and does not affect the original definition of the expression.

\subsection{BS to $S$ Link}
\subsubsection{Non-linear energy harvesting model}
We continue to use the complex non-linear harvesting model to determine the optimal number of modules.
This may necessitate incorporating more elements, i.e., in need of multiple modules, as we incorporate the information causality constraint in the system. The signal received from the BS to $S$ is:
\be
\begin{aligned}
\!\!\!\!s_{bs}(i^{'})=&\underbrace { \sqrt{p_{b}}h_{bs} s }_{\text {direct link}}+\underbrace {L(i^{'}) \mathbf {h_{rs}}\left ({ {p_{\mathrm {b}}}\mathbf {h_{br}} s +\mathbf {z}_{2} }\right)}_{{\text {RIS-aided link}}}+{z}_{1} \\=&\sqrt{p_{b}}\left ({ {h}_{bs}+ L(i^{'}) \mathbf {h_{rs}}\mathbf {h_{br}}}\right)s+ L(i^{'})\mathbf {h_{rs}} \mathbf {z}_{2} + {z}_{1}
\end{aligned}
\ee
where $p_b$ is the BS transmit power and $s$ represents the unit-power information signal. $\mathbf{z_2}$ is a complex-valued vector with dimensions $N\times 1$, representing the thermal noise generated by the RIS panel during the energy harvesting process from the BS RF signals. It follows a complex Gaussian distribution with mean 0 and variance $\sigma_2^2$. $z_1$ is the additive Gaussian channel noise, with a $\mathcal{CN}(0,\sigma_1^2)$ distribution.
$\mathbf {h_{rs}} \in \mathbb{C}^{1\times N}$ and $\mathbf {h_{br}} \in \mathbb{C}^{N\times 1}$ are the gains of RIS-$S$ and BS-RIS, respectively.
The harvested power from BS RF signals \cite{Ref_NEHModel} is expressed as follows:
\be
\begin{aligned}
\label{Eq_nl_ep}
p_{p,nl}^{b}(i^')=\frac{Y}{1+\exp(-y_1(p_{nl}^{b}(i^')-y_2))}
\end{aligned}
\ee

Note that (\ref{Eq_nl_ep}) is a non-linear model.
We define 
\be
\begin{aligned}
p_{nl}^{b}(i^')= \zeta p_b \frac{(h_{bs}+L(i^')  {h_3})^{2}}{  \sigma_1^2}
\end{aligned}
\ee
where ${h_3}=[\mathbf {h_{rs}}]_n [\mathbf {h_{rb}^{\dagger}}]_n  $.
$\zeta \in [0,1]$ is the power split coefficient indicating the portion of power assigned to the energy harvest unit.
$Y$, $y_1$, and $y_2$ define various parameters required for the wireless energy harvesting model.
The effect of resistance, capacitance, and circuit sensitivity is captured by $y_1$ and $y_2$.
Using (\ref{Eq_nl_ep}), non-linear energy harvested during $T(i^{''})$ is expressed as follows:
\be
\begin{aligned}
e_{nl}^{b}(i^')=\frac{YT(i^{'})}{1+\exp(-y_1(p_{nl}^{b}(i^')-y_2))}
\end{aligned}
\ee

\subsubsection{Data transmission from BS to $S$}
We calculate the data rate transmission from the BS to $S$ during  $T(i^{'})$ as follows:
\be
\begin{aligned}
\label{Rate_D2D}
R_{b}^{b}(i^')= W_2 \log _{2} \left ({ 1 + \frac { { (h_{bs}+L(i^') {h_3} })^{2}}{  \sigma_{m_1}^2} } \right)
\end{aligned}
\ee
where $W_2$ is the bandwidth of the carrier frequency. We define $\sigma_{m_1}^2=\sigma_2^2 L(i^{'})||\mathbf{h_{rs}} ||^2+\sigma_1^{2}$.

\subsection{D2D Communications}
The received signal from $S$ to $D$ during the second time slot, $T(i^{''})$ is:
\be
\begin{aligned}
\label{KUET}
\!\!\!& s_{sd}^{b}(i^{''})\!=\! \underbrace{\sqrt {\frac{e_{nl}^{b}(i^{'})}{T(i^{''})}}  h_{sd}s}_{\text{\textit{S-D} link}} \!+\! \underbrace{
 L(i^{''}) \mathbf{h_{rd}}  \bigg ( \!\!\sqrt{\frac{e_{nl}^{b}(i^{'})}{T(i^{''})}} p_b\mathbf{h_{sr}}\!+\!\mathbf{z_2} \!\bigg )}_{\text{RIS module-assisted link}}\!+\!z_1 \\
 & = \sqrt {\frac{e_{nl}^{b}(i^{'})}{T(i^{''})}} \bigg ( h_{sd}+L(i^{''})\mathbf{h_{rd}}\mathbf{h_{sr}} \bigg)s+L(i^{''}) \mathbf{h_{rd}}\mathbf{z_2}+z_1
\end{aligned}
\ee
where 
$\mathbf {h_{rs}} \in \mathbb{C}^{1\times N}$ and $\mathbf {h_{rd}} \in \mathbb{C}^{1\times N}$ are the gains of RIS-$S$ and RIS-$D$ links, respectively.
$\mathbf {h_{rs}^{\dagger}}$ is the conjugate transpose of $\mathbf {h_{sr}}$.
The data transmissions from the $S$ to $D$, followed by D2D communications protocol, during $T(i^{''})$ is expressed as follows:
\be
\begin{aligned}
\label{Rate_D2D}
R_{d}^{b}(i^{''})= W_3 \! \log _{2} \left ({ 1 + \frac {\frac{e_{nl}^{b}(i^')}{T(i^{''})} { ( h_{sd}+L(i^{''}){h_4} })^{2}}{\sigma_{m_2}^2} } \! \right)
\end{aligned}
\ee
where $W_3$ is bandwidth of the carrier frequency and ${h_4} =\mathbf {h_{rd}}\mathbf {h_{rs}^{\dagger}}$. We define $\sigma_{m_2}^2=\sigma_2^2 L(i^{''})||\mathbf{h_{rd}} ||^2+\sigma_1^{2}$. 

\subsection{Information Causality Constraint}
The BS sends information to $S$ in $T(i^{'})$ time, which is then
forwarded to $D$ in $T(i^{''})$.
The information causality constraint ensures that $S$ forwards to $D$ only the received information from the BS.
We express the information causality constraint as follows:
\be
\begin{aligned}
\label{Rate_D2D}
T(i^{'})R_b^{b}(i^{'}) \leq T(i^{''})R_d^{b}(i^{''})
\end{aligned}
\ee

\subsection{RIS Modules Energy Consumption} \label{sub_RISB}
We use a similar power consumption model from Section~\ref{Sec_III_E}.

\subsubsection{Power consumed by passive modules}
Recall that the RIS panel has $K_p$ passive modules in the proposed model, which consume energy due to control and switch circuits at the reflecting elements.
The power consumption due to the RIS passive module, used in both energy harvesting and D2D communications, is expressed as follows:
\be
\begin{aligned}
\label{eq_ac_powe}
p_{ss}^{b}(j)=K_p(j)k^* p_{sc}~\text{for}~j \in \{i^{'},i^{''}\}
\end{aligned}
\ee
where $p_{sc}$ is the power consumption due to switch and control circuits for each RIS passive element.

\subsubsection{Power consumed by active modules}
The RIS panel is equipped with $M_a$ active modules.
The power consumption of active modules includes a switch and control circuit power consumption and the power dissipation due to the loss in the DC circuit.
Power consumption during the energy harvesting phase is expressed as follows:
\be
\begin{aligned}
\label{eq_ac_powe}
p_{ct}^{b}(i^{'}) \! = \! M_a(i^{'})m^* (p_{sc} \! + \! p_{dc}) \! + \frac{{p_{out_1}}}{v}
\end{aligned}
\ee
where 
$p_{out_1} = a_m^2 p_b||\mathbf{h_{br}}||^2+\sigma_2^2$. 
The output power comprises two components: the energy consumed by the switch and control circuit, $p_{ss}$, and the DC biasing power consumption, $p_{dc}$.
The active module power consumption during the D2D communications phase is expressed as follows:
\be
\begin{aligned}
\label{eq_ac_powe}
p_{ct}^{b}(i^{''}) \! = \! M_a(i^{'})m^* (p_{sc} \! + \! p_{dc}) \! + \frac{{p_{out_2}}}{v}
\end{aligned}
\ee
where
$p_{out_2} = a_m^2 \frac{e_{nl}^{b}(i^{'})}{T(i^{''})}||\mathbf{h_{sr}}||^2+\sigma_2^2$.

\subsubsection{RIS modules energy consumption}
The RIS active and the passive modules' energy consumption during $T(i)$ is expressed as follows:
\be
\begin{aligned}
\label{eq_tot_powe}
e_{r}^{b}(i)= \sum_{j \in \{i^{'},i^{''}\}} \bigg [ p_{ss}^b(j)+p_{ct}^b(j)\bigg]
\end{aligned}
\ee

\subsection{Optimization Problem}
We formulate the optimization problem minimizing overall RIS modules' energy consumption subject to the information causality constraint, non-linear energy harvesting model, and the number of RIS modules.
The optimization problem, \textbf{P.2} is:
\begin{subequations}\label{eq_op_5}
\begin{align}\label{eq_ob_5}\
\textbf{P.2}~&{\mathop {\min }\limits_{M_a,K_p,p_{b},T(i{'}),T(i^{''})}
}\ {\!\!\!\! \sum_{j \in \{i^{'},i^{''}\}}\!\!\!\!T(j)  \bigg [p_{ss}^{b}(j) +p_{ct}^{b}(j) \bigg]}\\
&\text{s.t.}\ M_am^*+K_pk^* \leq N \label{eq_opt5_c1}\\
& M_a(j) \leq M_a,~~K_p(j) \leq K_p~\text{for}~ j \in \{i^{'},i^{''}\} \\
& T(i^{'})+T(i^{''}) = T(i) \label{eq_opt5_c10}\\ 
& e_m \leq \frac{YT(i^{'})}{1+\exp(-y_1(p_{nl}^{b}(i^')-y_2))} \label{eq_opt5_c2}\\
&\!\begin{aligned} & \! T(i^{''})W_3 \log _{2} \left ({ 1 + \frac { \frac{e_{nl}^{b}(i^')}{T(i^{''})} ({h_{sd}+ L(i^{''}){h_4} })^{2}}{\sigma_{m_2}^2} } \! \right) \geq \\ & T(i^{'})W_2\log _{2} \left ({ 1 + \frac { p_b({ h_{bs}+L(i^{'}){h_3} })^{2}}{  \sigma_{m_1}^2} } \right)
    \end{aligned} \label{eq_opt5_c4}
\end{align}
\end{subequations}

The objective function minimizing the RIS modules' energy consumption during $T(i)$ is described in (\ref{eq_ob_5}).
The total number of elements combined in active and passive modules is shown in (\ref{eq_opt5_c1}), where $N$ is the maximum number of RIS elements used to design the modules.
non-linear constraint in (\ref{eq_opt5_c2}) defines the harvested energy from BS RF signals by $S$ to be no less than a threshold, $e_m$.
The information causality constraint is defined in (\ref{eq_opt5_c4}).
(\ref{eq_op_5}) is a non-convex MINLP due to the coupling of optimizing variables, making it very difficult to solve optimally. 
Section~\ref{OAPM} outlines a heuristic solution approach to tackle the challenge.

\section{Optimal Number of RIS Modules} \label{OAPM}
To simplify the complexity of (\ref{eq_op_5}), we break it into three smaller sub-problems and solve them in iteration. First, we optimize the number of RIS modules, $M_a$ and $K_p$, given the power budget, $p_b$, and the time slots, $T(i^{'})$ and $T(i^{''})$. Next, using the optimal values of $M_a^*$ and $K_p^*$, we optimize $T(i^{'})$ and $T(i^{''})$. Finally, we use these optimal values to find the optimal BS power, $p_b^*$. We use feasible points as an approximation to make the problem more tractable. The two sub-problems will be examined in the following sub-sections, solved iteratively.

\subsection{RIS Optimal Modules, $M_a^*$ and $K_p^*$} \label{Sub_iiB}
Given $T(i^{'})$, $T(i^{''})$, and $p_b$,  we optimize the number of RIS modules, $M_a$, and $K_p$.
We, therefore, reformulate (\ref{eq_op_5}) as follows:
\begin{subequations}\label{eq_op_51}
\begin{align}\label{eq_ob_51}\
&{\textbf{P.2A}~ \mathop {\min }\limits_{M_a,K_p}
}\ { \sum_{j \in \{i^{'},i^{''}\}}T(j) \bigg [p_{ss}^{b}(j) +p_{ct}^{b}(j) \bigg]}\\
&\text{s.t.}\ M_am^*+K_pk^* \leq N \label{eq_opt51_c1}\\
& M_a(j) \leq M_a,~~K_p(j) \leq K_p~\text{for}~ j \in \{i^{'},i^{''}\} \label{eq_opt51_c1....} \\
& e_m \leq \frac{YT(i^{'})}{1+\exp(-y_1(p_{nl}^{b}(i^')-y_2))} \label{eq_opt51_c2} \\
&\!\begin{aligned} & \! T(i^{''})W_3 \log _{2} \left ({ 1 + \frac { \frac{e_{nl}^{b}(i^')}{T(i^{''})} ({h_{sd}+ L(i^{''}){h_4} })^{2}}{\sigma_{m_2}^2} } \! \right) \geq \\ & T(i^{'})W_2\log _{2} \left ({ 1 + \frac { p_b({ h_{bs}+L(i^{'}){h_3} })^{2}}{  \sigma_{m_1}^2} } \right)
    \end{aligned}  \label{eq_opt51_c4} 
\end{align}
\end{subequations}

After some mathematical manipulations, we reformulate the non-convex energy harvesting constraint in (\ref{eq_opt51_c2}) by taking the exponential term out as follows: 
\be
\begin{aligned}
\label{Eq_51_C3_N1}
~~& \! \! \! f_{10} \! \leq \! h_{bs}^2+2h_{bs}h_3\bigg( m^*\sum_{m=1}^{M_a(i^{'})}a_m +K_p(i^{'})k^*\bigg) + \\
&\bigg({m^*}^2 \sum_{m=1}^{M_a^f(i^{'})}a_m \sum_{m=1}^{M_a(i^{'})}a_m  + 2 m^*  k_p(i^{'})k^* \sum_{m=1}^{M_a^f(i^{'})}a_m  \\
& K_p^f(i^{'})K_p(i^{'})k^* \bigg) h_{3}^2
\end{aligned}
\ee
where 
\be
\begin{aligned}
f_{10}=\sigma_{1}^2\frac{y_1y_2-\ln\bigg(\frac{YT(i^')}{e_m}-1\bigg)}{y_1\zeta p_b}
\end{aligned}
\ee

We decouple the optimizing variables in $p_{ct}^{b}(i^{''})$ to solve the objective function in (\ref{eq_ob_51}).
We reformulate $p_{ct}^{b}(i^{''})$ as follows:
\be
\begin{aligned}
\label{eq_ac_powe}
\!\!p_{ct}^{b^f}(i^{''}) \!  = \! M_a(i^{'})m^* \! (p_{sc} \! + \! p_{dc}) \! +\! \frac{{  \frac{a_m^2 ||\mathbf{h_{sr}}||^2 }{T(i^{''})} e_{nl}^{b^f}(i^{'}) +\sigma_2^2}}{v} \!\!
\end{aligned}
\ee
where 
\be
\begin{aligned}
\label{eq_ac_powefghfgh}
\!\!\!e_{nl}^{b^f}(i^')=\frac{YT(i^{'})}{1+\exp\bigg(-y_1\bigg(\zeta p_b \frac{(h_{bs}+L^f(i^')  {h_3})^{2}}{  \sigma_1^2}-y_2\bigg)\bigg)}
\end{aligned}
\ee
and 
\be
\begin{aligned}
\label{bolbona}
L^f(i^{'})= m^* \sum_{m=1}^{M_a^f(i^{'})}a_m +K_p^f(i^{'})k^*
\end{aligned}
\ee

Using (\ref{eq_ac_powefghfgh}), we redefine the information causality constraint as follows:
\be
\begin{aligned} 
& \log _{2} \left ({ 1 +  f_{12} ({h_{sd}+ L(i^{''}){h_4} })^{2} } \! \right) \geq 
\\ & \log _{2} \left ({ 1 +  f_{13}({ h_{bs}+L(i^{'}){h_3} })^{2} } \right)^{f_{11} }
\end{aligned}
\ee
where $f_{11}=\frac{T(i^{''})W_3}{T(i^{'})W_2}$, $f_{12}=\frac{e_{nl}^{b^f}(i^{'})}{\sigma_{m_2}^2 T(i^{''})}$, and $f_{13}=\frac{p_b}{\sigma_{m_1}^2}$.
Using (\ref{bolbona}) and taking $\log$ term out, the following expression can be found:
\be
\begin{aligned} 
\label{Kothin}
\frac{\sqrt{\frac{{ f_{13}({ h_{bs}+L^f(i^{'}){h_3} })^{2} }^{f_{11} }}{f_{12}}} - h_{sd} }{h_4}\leq {{ L(i^{''}) } } 
\end{aligned}
\ee

The reformulated problem from (\ref{eq_op_51}) is expressed as follows:
\begin{align} \label{eq_op_511}
& \underset{M_a^*,K_p^*}{\text{min}} \ T(i^{'})p_{ct}^{b}(i^{'})+T(i^{''})p_{ct}^{b^f}(i^{''}) +\!\!\!\!\!\!\sum_{j \in \{i^{'},i^{''}\}} \!\!\!\!\! T(j)  p_{ss}^{b}(j)  \\
&\text{s.t.}\ (\ref{eq_opt51_c1}), (\ref{eq_opt51_c1....}),(\ref{Eq_51_C3_N1}), (\ref{Kothin}) \nonumber
\end{align}

Note (\ref{eq_op_511}) is a convex approximation to the problem of (\ref{eq_op_51}).
Hence, the solutions, $M_a^*$ and $K_p^*$ are obtained from (\ref{eq_op_511}), given that $T(i^{'})$, $T(i^{''})$, and $p_b$ are known.

\subsection{Optimal Time Slots, $T^*(i{'})$ and $T^*(i^{''})$ }
In this subsection, we retain $M_a^*$ and $K_p^*$ to find optimal $T^*(i{'})$ and $T^*(i^{''})$. We reformulate the optimization problem from (\ref{eq_op_5}) as follows:
\begin{subequations}\label{eq_s2op_5}
\begin{align}\label{eq_s2ob_5}\
&{\textbf{P.2B}~\mathop {\min }\limits_{T(i{'}),T(i^{''})}
}\ { \sum_{j \in \{i^{'},i^{''}\}}T(j) \bigg [p_{ss}^{b}(j) +p_{ct}^{b}(j) \bigg]}\\
&\text{s.t.}\  T(i^{'})+T(i^{''})=T(i) \label{eq_s2opt5_c10}\\ 
& e_m \leq \frac{YT(i^{'})}{1+\exp(-y_1(p_{nl}^{b}(i^')-y_2))} \label{eq_s2opt5_c2}\\
&\!\begin{aligned} & \! T(i^{''})W_3 \log _{2} \left ({ 1 + \frac { \frac{e_{nl}^{b}(i^')}{T(i^{''})} ({h_{sd}+ L(i^{''}){h_4} })^{2}}{\sigma_{m_2}^2} } \! \right) \geq \\ & T(i^{'})W_2\log _{2} \left ({ 1 + \frac { p_b({ h_{bs}+L(i^{'}){h_3} })^{2}}{  \sigma_{m_1}^2} } \right)
    \end{aligned} \label{eq_s2opt5_c4}
\end{align}
\end{subequations}

We reformulate (\ref{eq_s2opt5_c4}) as follows:
\be
\begin{aligned} 
\label{Eq_77}
\! T(i^{''}) \log _{2} \left ({ 1 + \frac {f_{14} }{T(i^{''})} } \! \right) \geq  T(i^{'}) f_{15}
\end{aligned}
\ee
where 
\be
\begin{aligned} 
f_{14}=\frac{e_{nl}^{b^f}(i^')}{\sigma_{m_2}^2} ({h_{sd}+ L(i^{''}){h_4} })^{2}
\end{aligned}
\ee
\be
\begin{aligned}
e_{nl}^{b^f}(i^')=\frac{YT^f(i^{'})}{1+\exp(-y_1(p_{nl}^{b}(i^')-y_2))}
\end{aligned}
\ee
and 
\be
\begin{aligned} 
f_{15}=\frac{W_2}{W_3}\log _{2} \left ({ 1 + \frac { p_b({ h_{bs}+L(i^{'}){h_3} })^{2}}{  \sigma_{m_1}^2} } \right)
\end{aligned}
\ee

It's generally difficult to make (\ref{Eq_77}) convex. 
To find a convex lower bound for (\ref{Eq_77}), using a convex relaxation technique, such as the log-sum-exp (LSE) \cite{neely2006energy} can be used to approximate as follows:
\be
\begin{aligned}
\label{Eq_78}
T(i^{''}) \log _{2} \left ({ e + \frac {f_{14} }{T(i^{''})} } \! \right) \geq  T(i^{'}) f_{15}
\end{aligned}
\ee

Note that this approximation in (\ref{Eq_78}) is a convex function, so it can be optimized using the LSE convex optimization technique. 
The reformulated problem from (\ref{eq_s2op_5}) is expressed as follows:
\begin{align} \label{eq_s2op_dsfd5}
& \underset{T^*(i^{'}),T^*(i^{''})}{\text{min}} \ \!\!\!\!\!\!\!\!T(i^{'})p_{ct}^{b}(i^{'})+T(i^{''})p_{ct}^{b^f}(i^{''}) \!+\!\!\!\!\!\!\!\!\sum_{j \in \{i^{'},i^{''}\}} \!\!\!\!\! T(j)  p_{ss}^{b}(j)  \\
&\text{s.t.}\ (\ref{eq_s2opt5_c10}), (\ref{eq_s2opt5_c2}),(\ref{Eq_78}) \nonumber
\end{align}

Note (\ref{eq_s2op_dsfd5}) is a convex approximation to the problem of (\ref{eq_s2op_5}).
Hence, the solutions, $T^*(i^{'})$ and $T^*(i^{''})$ are obtained from (\ref{eq_op_511}), when $M_a^*$, $K_p^*$, and $p_b$ are available.
    
\subsection{Optimal BS Power Budget, $p_b^*$}
In this sub-section, we use optimal $M_a^*$, $K_p^*$, $T^*(i^{'})$ and $T^*(i^{''})$ from Section~\ref{Sub_iiB} to achieve the optimal BS power budget, $p_b^*$.
We reformulate the optimization problem from (\ref{eq_op_5}) as follows:
\begin{subequations}\label{eq_op_53}
\begin{align}\label{eq_ob_53}\
&{\textbf{P.2C}~\mathop {\min }\limits_{p_{b}}
}\ { \sum_{j \in \{i^{'},i^{''}\}}T(j) \bigg [p_{ss}^{b}(j) +p_{ct}^{b}(j) \bigg]}\\
&\text{s.t.}\ e_m \leq \frac{YT(i^{'})}{1+\exp(-y_1(p_{nl}^{b}(i^')-y_2))} \label{eq_opt53.}\\
&\!\begin{aligned} & \! T(i^{''})W_3 \log _{2} \left ({ 1 + \frac { \frac{e_{nl}^{b}(i^')}{T(i^{''})} ({h_{sd}+ L(i^{''}){h_4} })^{2}}{\sigma_{m_2}^2} } \! \right) \geq \\ & T(i^{'})W_2\log _{2} \left ({ 1 + \frac { p_b({ h_{bs}+L(i^{'}){h_3} })^{2}}{  \sigma_{m_1}^2} } \right)
    \end{aligned} \label{eq_opt53}
\end{align}
\end{subequations}

The exponential term of (\ref{eq_opt53.}) is dealt with as follows:
\be
\begin{aligned}
\label{eq_op_53_n}
p_b \geq  \frac{  \sigma_{m_1}^2\bigg (y_2-\frac{\ln(\frac{YT(i^')}{e_m}-1)}{y_1}\bigg) }{\zeta (h_{bs}+L(i^') {h_3})^2} 
\end{aligned}
\ee

We rewrite (\ref{eq_opt53}) as follows:

\be
\begin{aligned}
\label{almost}
f_{18} \geq \log _{2} \left ({ 1 + f_{17} p_b } \right)
\end{aligned}
\ee
where
\be
\begin{aligned}
f_{18}=\frac{T(i^{''})W_3}{T(i^{'})W_2} \log _{2} \left ({ 1 + e_{nl}^{b^f}(i^') \frac {  ({h_{sd}+ L(i^{''}){h_4} })^{2}}{T(i^{''})\sigma_{m_2}^2} } \! \right)
\end{aligned}
\ee
\be
\begin{aligned}
f_{17}=\frac { ({ h_{bs}+L(i^{'}){h_3} })^{2}}{  \sigma_{m_1}^2}
\end{aligned}
\ee
and
\be
\begin{aligned}
e_{nl}^{b^f}(i^')=\frac{YT(i^{'})}{1+\exp \bigg (-y_1 \bigg (\zeta p_b^f \frac{(h_{bs}+L(i^')  {h_3})^{2}}{  \sigma_1^2}-y_2 \bigg ) \bigg )}
\end{aligned}
\ee

Note that (\ref{almost}) is a concave function, meaning its second derivative is negative. To make it a convex function, we can take its negative, as the negative of a concave function is a convex function.
Therefore, the following function is convex:
\be
\begin{aligned}
&-\log_2(1 + p_b f_{17}) = -\log_2(1 + f_{17}p_b) \\
& = \log_2((1 + f_{17}p_b)^{-1})  = \log_2(2^{-f_{18}} - f_{17}p_b)
\end{aligned}
\ee

And the inequality can be written as:
$\log_2(1 + f_{17}p_b) \leq f_{18}$, which is equivalent to $\log_2(2^{-f_{18}} - f_{17}p_b) \geq 0$.

The inequality is only true for values of $f_{17}p_b$ that satisfy the constraint such that 
\be
\begin{aligned}
\label{ok}
\bigg ( 2^{-f_{18}} - f_{17}p_b \bigg ) > 0
\end{aligned}
\ee
In order to make (\ref{eq_ob_53}), we approximate (\ref{eq_ac_powe}) to be convex as follows:
\be
\begin{aligned}
\label{eq_RISsdsd}
p_{ct}^f(i^{''}) \! = \! m(i^{''}) (p_{sc} \! + \! p_{dc}) \! + \frac{a_m^2 \frac{e_{nl}^{b^f}(i^')}{T(i^{''})}|| \mathbf{h_{sr}}||^2+\sigma_2^2}{v}
\end{aligned}
\ee
where 
\be
\begin{aligned}
e_{nl}^{b^f}(i^')=\frac{YT(i^{'})}{1+\exp(-y_1(\zeta p_b^f \frac{(h_{bs}+L(i^')  {h_3})^{2}}{  \sigma_1^2}-y_2))}
\end{aligned}
\ee

We rewrite the optimization problem as follows:
\begin{align} \label{eq_op_53}
& \underset{p_{b}^*}{\text{min~~}} {T(i^{'})p_{ct}^{b}(i^{'}) +T(i^{''})p_{ct}^{b^f}(i^{''}) + \!\!\!\!\sum_{j \in \{i^{'},i^{''}\}} \!\!\!\! T(j)  p_{ss}^{b}(j) } \\
&\text{s.t.}\ (\ref{eq_op_53_n}), (\ref{ok}) \nonumber
\end{align}

Note (\ref{eq_op_53}) is a convex function, it can be solved using any standard optimization toolbox, which provides a range of algorithms to solve convex optimization problems. Therefore, the problem of optimizing (\ref{eq_op_53}) can be tackled using well-established and efficient optimization techniques, making the proposed solution practical and scalable.

\subsection{Proposed Algorithm}
The proposed solution involves an iterative process to solve three equations, namely (\ref{eq_op_511}), (\ref{eq_s2op_dsfd5}), and (\ref{eq_op_53}), until convergence is achieved. The algorithm for this solution can be summarized as follows:
\begin{algorithm}
\caption{}
\label{alg:algorithm_sum}
\begin{algorithmic}[1]
\State $\bold{Initialization}$: $M_a^f$, $K_p^f$, $p_b^f$,  $T^f(i^{'})$, $T(i)$, $l=0$, and $\epsilon$ to some reasonable random values
\State $\bold {Optimization}$:
\State Set $l \leftarrow l+1$ 
\Repeat
\State Find the number of RIS active and passive modules to obtain optimal $M_a^*$ and $K_p^*$ using (\ref{eq_op_511})
\State Set $M_a^f=M_a^*$ and $K_p^f=K_p^*$
\State Using $M_a^*$ and $K_p^*$, find obtain optimal, $T^*(i^{'})$ and $T^*(i^{''})$ using (\ref{eq_s2op_dsfd5})
\State Set $T^f(i^{'})=T^*(i^{'})$
\State Using $M_a^*$, $K_p^*$, $T^*(i^{'})$ and $T^*(i^{''})$, find the optimal transmit power, $p_b^*$, using (\ref{eq_op_53})
\State Set  $p_b^f=p_b^*$
\Until{Find the optimal value, such as $m^*$, $k^*$, $T^*(i^{'})$, $T^*(i^{''})$, and $p_b^*$ such that $|x(l)|-|x(l-1)|< \epsilon$, where $x=M_a^*,K_p^*,T^*(i^{'}),T^*(i^{''}),p_b^*$} 
\end{algorithmic}
\end{algorithm}
\begin{figure}[H]
\centering
\includegraphics[width=2.7in]{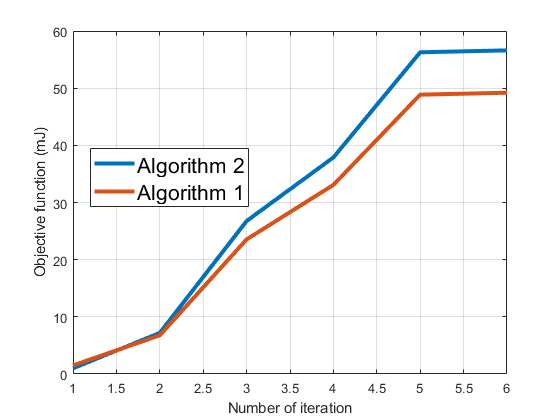}
\caption{Convergence of Algorithms}
\label{Fig_Convergence}
\end{figure}
\begin{figure}[H]
\centering
\includegraphics[width=2.7in]{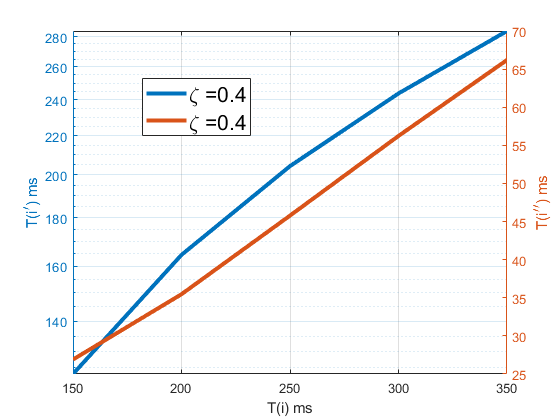}
\caption{Time management}
\label{Fig_Time}
\end{figure}

\section{Results} \label{RA12}
This section presents the simulation setup of the proposed model, including specific parameter details, and outlines the results obtained from the simulation.
Our investigation focuses on stationary nodes, where their distances determine the gain between nodes. The channel gain is based on Rician fading, a commonly used model in wireless communication that accounts for both line-of-sight and non-line-of-sight components of signal propagation. The algorithms are simulated using a specific set of parameters, which include the thermal noise $-121.45$ dBm, $e_m=3$ mJ, $a_m \in \{2.5, 3.0, 3.5\}$, $p_{sc}=-10$ dBm, $p_{dc}=-5$ dBm.
We use $180$ KHz as the bandwidth.
The energy efficiency factor and time are set as $\zeta \in \{ 0.20,0.30,0.40\}$ and $T(i) \in \{150, 200,250,300,350\}$ ms.  
We assume the value of $Y$ to be 1, indicating that the maximum amount of power that can be charged is 1 W.
The RIS panel, a key component of the proposed system, is designed with a minimum of 150 and a maximum of 850 elements. We also set $R_b^t=R_d^t=0.1$ bps/Hz. To solve the optimization performance, we use the GAMS tool, which is a high-level modeling system for mathematical programming and optimization. 
In Fig.~\ref{Fig_Convergence}, the objective function increases within the first five iterations, and after that, it stabilizes and changes very little. This indicates the convergence of the algorithms, and the results produced by the algorithms are reliable.
The fast convergence of Algorithms is a desirable feature, as it indicates that these algorithms can solve the problem efficiently and quickly. 
\begin{figure}[H]
\centering
\includegraphics[width=2.7in]{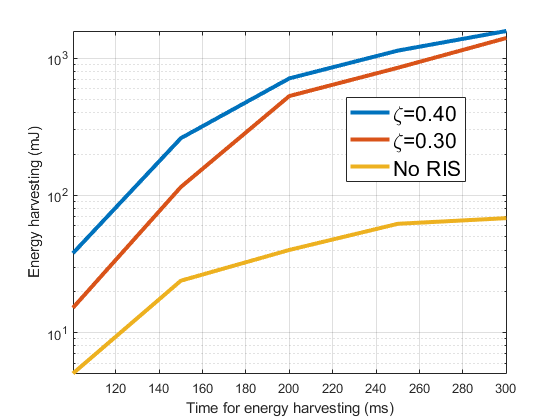}
\caption{Energy harvesting varying over time}
\label{Fig_NonLinearEH}
\end{figure}
\begin{figure}[H]
\centering
\includegraphics[width=2.7in]{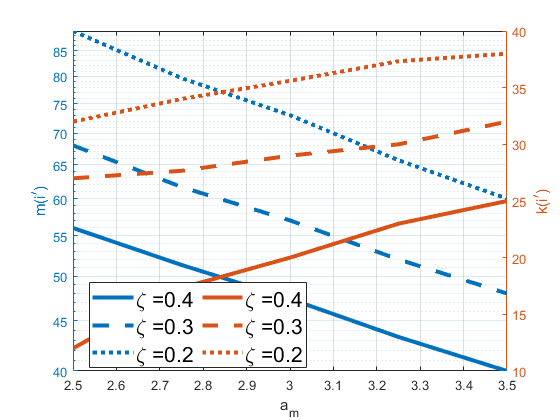}
\caption{Elements during energy harvesting}
\label{Fig_m1}
\end{figure}

\begin{figure}[H]
\centering
\includegraphics[width=2.7in]{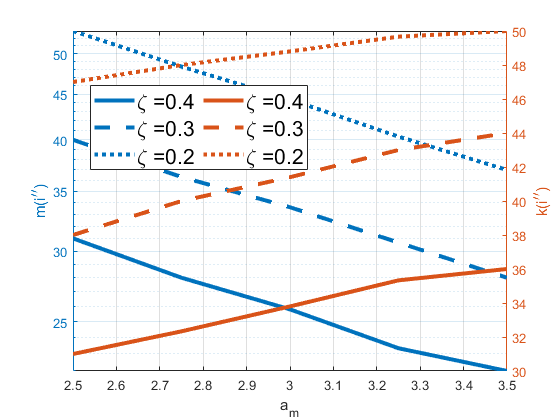}
\caption{Elements during D2D communications}
\label{Fig_m2}
\end{figure}

Fig.~\ref{Fig_Time} presents a comparison of different $T(i)$ values and their corresponding optimal times $T(i')$ and $T(i")$ for energy harvesting from the BS RF signals and D2D communications, respectively.
The finding suggests that the time required for energy harvesting is much higher than the D2D communications time in a given time frame. 
When the efficiency factor, $\zeta$, is high, a higher amount of energy can be harvested, and more information can be transmitted from $S$ to $D$, which results in an overall increase in time needed for both cases.
The energy harvesting varying over time is shown in Fig.~\ref{Fig_NonLinearEH}.
The higher efficiency factor guarantees higher energy harvesting in both algorithms.
Specifically, the non-linear energy harvesting model yields more energy.
For example, with $\zeta=0.40$, compared to a lower efficiency factor, such as $\zeta=0.30$. Notably, the absence of the RIS panel in the networks leads to poor energy harvesting performance.
\begin{figure}[H]
\centering
\includegraphics[width=2.7in]{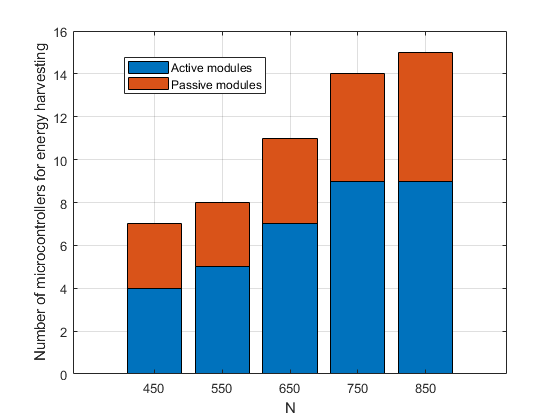}
\caption{Number of microcontrollers  used in energy harvesting}
\label{Fig_microcontroller_EH}
\end{figure}
\begin{figure}[H]
\centering
\includegraphics[width=2.7in]{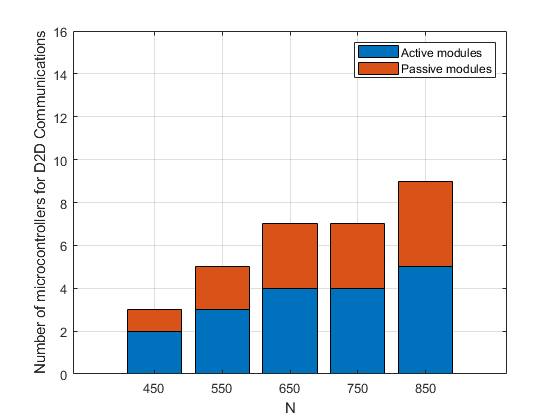}
\caption{Number of microcontrollers for D2D communications}
\label{Fig_microcontroller_D2D}
\end{figure}

In Fig.~\ref{Fig_m1} and Fig.~\ref{Fig_m2}, we present the variation in the optimal number of active and passive elements involved in the energy harvesting process and D2D communications, respectively. The optimal number of elements is captured against the amplification factor, $a_m$, while the change to different values of $\zeta$ is analyzed. It is observed that the number of active elements decreases with an increase in $a_m$ for both cases. 
For instance, the active elements utilize 60\% of the overall RIS elements (Fig.~\ref{Fig_m1}), while in Fig.~\ref{Fig_m2}, they utilize 34\% of the overall RIS elements.
The difference in the number of active elements used can be attributed to the complexity of the energy harvesting and D2D communication processes.
The optimal number of passive elements for the energy harvesting and D2D communications phases are also presented in Fig.~\ref{Fig_m1} and Fig.~\ref{Fig_m2}. It is observed that the number of optimal passive elements slightly increases for a higher number of RIS elements. Specifically, for energy harvesting, the passive elements use only 21\% of the overall RIS elements, while for D2D communications, they use 30\% of the RIS elements. This difference in the percentage of passive elements used can be attributed to the role of active and passive elements in each phase. In the energy harvesting phase, the active elements with higher $a_m$ are more important for achieving higher energy gain, whereas, in D2D communications, passive elements play a bigger role in reflecting and directing the signals between the communicating nodes.
\begin{figure*}
\begin{subfigure}[H]{0.39\textwidth}
\centering
\includegraphics[width=\textwidth]{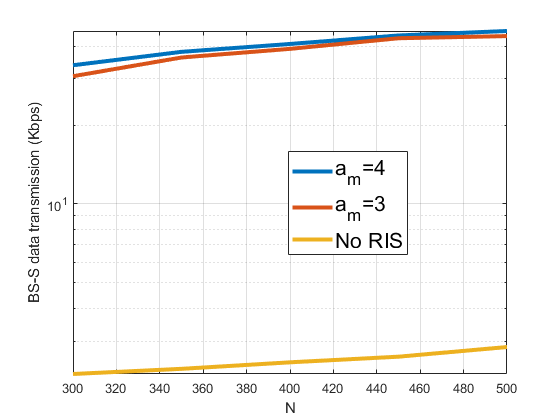}
\caption{Data transmission from BS to $S$}
\label{Fig_BS_Rate}
\end{subfigure}
\hfill
\begin{subfigure}[H]{0.39\textwidth}
\centering
\includegraphics[width=\textwidth]{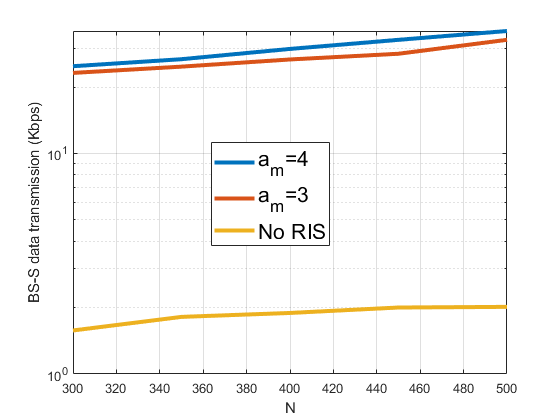}
\caption{Data transmission from $S$ to $D$}
\label{Fig_D2D_Rate}
\end{subfigure}
\caption{Information causality constraint}
\label{Fig_Rate}
\end{figure*}

Fig.\ref{Fig_microcontroller_EH} and Fig.\ref{Fig_microcontroller_D2D} show the variation in the number of microcontrollers for different values of $N$. It is important to note that RIS modules are designed using the optimal combination of active and passive elements, denoted as $m^*$ and $k^*$, respectively, as determined by Algorithm 1. Each microcontroller is responsible for controlling one RIS module. Thus, in this case, the total number of RIS modules equals the total number of microcontrollers, equal to $m^*=k^*=1$ microcontroller.
The figures show the number of microcontrollers used during the energy harvesting and D2D communications phases. D2D communications requires fewer microcontrollers than the non-linear energy harvesting model. However, as $N$ increases, RIS modules also increase. The microcontroller utilization for RIS active and passive modules during energy harvesting increases to 56
The figures also show the number of microcontrollers used during D2D communications from the source $S$ to the destination $D$ phase. When $N$ is 450 and 850, the number of microcontrollers for active elements changes from 2 to 5 with a 60\% increment, while the number of microcontrollers for passive elements changes from 1 to 4 with a 75\% increment.
Comparing Figures~\ref{Fig_microcontroller_EH} and~\ref{Fig_microcontroller_D2D}, we can conclude that the RIS modules concept is particularly suitable when the number of elements is large.

Fig.~\ref{Fig_BS_Rate} and Fig.~\ref{Fig_D2D_Rate} show the data rate transmission using RIS modules. To demonstrate the performance, we vary $a_m$ for different numbers of RIS elements, such as $N \in \{300, 500\}$. We compare the RIS-assisted data rate with no RIS panel, and the proposed model shows an enhanced transmission rate of about 100\% compared to the network without the RIS panel. It is reasonable to claim that the higher number of RIS modules results in a higher data rate, as demonstrated in Fig.~\ref{Fig_Rate}. Comparing Fig.~\ref{Fig_BS_Rate} and Fig.~\ref{Fig_D2D_Rate}, we observe that the rate from $S$ to $D$ is higher or equal to the rate from BS to $S$, indicating that the information causality constraint is successfully applied to the network. This claim also holds when the system does not adopt any RIS panel, as the information causality constraint is independent of the RIS panel.

\section{Conclusions} \label{End}
In this paper, we investigated the impact and control of RIS active and passive elements to enhance the performance of wireless communications for b-IoT systems. Active elements in the RIS panel consume energy for signal reflection and amplification, while passive elements consume energy only for signal reflection. RISs require the use of microcontrollers for controlling purposes. 
However, using a larger number of elements with microcontroller-based intelligent controls leads to increased complexity. To address this challenge, we study the trade-off between the number of active and passive elements and the number of microcontrollers required to control them.
We proposed a unique and innovative approach, referred to as "Module," for the first time in the literature. The approach employs a single microcontroller to control a single module that is defined as either active or passive, with each module containing an optimal number of active or passive elements. The optimal size of a module is determined using a non-linear energy harvesting model, in which a b-IoT sensor harvests energy from the nearby BS RF signals. Once the optimal size is obtained, we achieve the optimal number of modules, i.e., the optimal number of microcontrollers required for efficient RIS panel control. With the harvested energy from the BS RF signals, the b-IoT sensor transmits data from the BS to other IoT sensors while enforcing an information causality constraint in RIS module-assisted IoT systems.
We minimize the energy consumption of the overall RIS modules and obtain the optimal number of active and passive modules, i.e., the number of microcontrollers, subject to non-linear energy harvesting model and information causality constraint data transmission constraints. We formulate the optimization problem as a non-convex MINLP problem, which is relaxed and solved using an iterative algorithm. Finally, simulation results show that the RIS panel with active and passive modules improves the performance of the b-IoT system compared to baseline schemes, such as no RIS panel, by optimizing the number of microcontrollers for a larger number of RIS elements.

\bibliographystyle{IEEEtran}
\bibliography{IEEEabrv,Reference/mybib}

\end{document}